\documentclass[
10pt, 
a4paper, 
oneside, 
headinclude,footinclude, 
BCOR5mm, 
]{scrartcl}

\usepackage[
nochapters, 
beramono, 
eulermath,
pdfspacing, 
dottedtoc 
]{classicthesis} 

\usepackage{arsclassica} 

\usepackage[T1]{fontenc} 

\usepackage[utf8]{inputenc} 

\usepackage{graphicx} 
\graphicspath{{Figures/}} 

\usepackage{enumitem} 

\usepackage{lipsum} 

\usepackage{subfig} 

\usepackage{amsmath,amssymb,amsthm} 

\usepackage{varioref} 

\theoremstyle{definition} 

\theoremstyle{plain} 

\theoremstyle{remark} 

\hypersetup{
colorlinks=true, breaklinks=true, bookmarks=true,bookmarksnumbered,
urlcolor=webbrown, linkcolor=RoyalBlue, citecolor=webgreen, 
pdftitle={}, 
pdfauthor={\textcopyright}, 
pdfsubject={}, 
pdfkeywords={}, 
pdfcreator={pdfLaTeX}, 
pdfproducer={LaTeX with hyperref and ClassicThesis} 
} 

\usepackage[a4paper, total={160mm,257mm}]{geometry}

\usepackage{subfig}
\usepackage{tabularx}
\usepackage{multicol}
\usepackage{multirow}
\usepackage{makecell}
\usepackage{enumitem}
\usepackage{pifont}
\usepackage{soul}
\usepackage{cite}
\usepackage{amsmath,amssymb,amsfonts}
\usepackage{algorithmic}
\usepackage{graphicx}
\usepackage{textcomp}
\usepackage{xcolor}
\usepackage{url}
\usepackage{tikz}
\usepackage[absolute]{textpos}
\usepackage{tcolorbox}

\usepackage{listings}
\usepackage{authblk}
\providecommand{\keywords}[1]
{
  \small	
  \textbf{\textit{Keywords---}} #1
}

\definecolor{armygreen}{rgb}{0.29, 0.33, 0.13}

\title{\normalfont\spacedallcaps{Berserker: ASN.1-based Fuzzing of Radio Resource Control Protocol for \\4G and 5G}}

\author{
  \spacedlowsmallcaps{Srinath Potnuru\textsuperscript{1,*} \& Prajwol Kumar Nakarmi\textsuperscript{2}} \\
  \small{\textsuperscript{1}Department of Computer Science, KTH Royal Institute of Technology, Sweden\\\emph{potnuru@kth.se}}
  \\ 
  \small{\textsuperscript{2}Business Area Networks, Ericsson, Sweden\\\emph{prajwol.kumar.nakarmi@ericsson.com}}

} 

\date{2021} 

\begin{document}


\renewcommand{\sectionmark}[1]{\markright{\spacedlowsmallcaps{#1}}} 
\lehead{\mbox{\llap{\small\thepage\kern1em\color{halfgray} \vline}\color{halfgray}\hspace{0.5em}\rightmark\hfil}} 

\pagestyle{scrheadings} 


\maketitle

\begin{textblock}{12.5}(2,0.2)
  \begin{tcolorbox} [colback=white,colframe=black, sharp corners, boxsep=0mm, boxrule=0.2mm]
    \footnotesize The paper is published in the 17th International Conference on Wireless and Mobile Computing, Networking and Communications (WiMob 2021) October 11–13, 2021, Virtual. ISBN: 978-1-6654-2854-5. DOI: 10.1109/WiMob52687.2021.9606317 \textcopyright 2021 IEEE. Personal use of this material is permitted. Permission from IEEE must be obtained for all other uses, in any current or future media, including reprinting/republishing this material for advertising or promotional purposes, creating new collective works, for resale or redistribution to servers or lists, or reuse of any copyrighted component of this work in other works.
  \end{tcolorbox}
\end{textblock}

\section*{abstract}
  Telecom networks together with mobile phones must be rigorously tested for robustness against vulnerabilities in order to guarantee availability. RRC protocol is responsible for the management of radio resources and is among the most important telecom protocols whose extensive testing is warranted. To that end, we present a novel RRC fuzzer, called \textit{Berserker}, for 4G and 5G. Berserker's novelty comes from being backward and forward compatible to any version of 4G and 5G RRC technical specifications. It is based on RRC message format definitions in ASN.1 and additionally covers fuzz testing of another protocol, called NAS, tunneled in RRC. Berserker uses concrete implementations of telecom protocol stack and is unaffected by lower layer protocol handlings like encryption and segmentation. It is also capable of evading size and type constraints in RRC message format definitions. Berserker discovered two previously unknown serious vulnerabilities in srsLTE -- one of which also affects openLTE -- confirming its applicability to telecom robustness.

\keywords{
  Fuzzing, security, RRC, NAS, ASN.1, 4G, 5G
}

\newcommand\blfootnote[1]{%
  \begingroup
  \renewcommand\thefootnote{}\footnote{#1}%
  \addtocounter{footnote}{-1}%
  \endgroup
}

\blfootnote{\textsuperscript{*}The work was a part of the first author's Master thesis at Ericsson.}

\section{Introduction} \label{introduction}

\textbf{Context.} Telecom networks enable not only mobile broadband connection, but also (increasingly) industry automation and critical communications \cite{Moteff2003CriticalIW, Statisti27:online}. They are required to have 99.999\% (five 9s) availability \cite{five_nine, Gray1991HighavailabilityCS}, which is less than six minutes of downtime in a year. To ensure high availability, telecom networks must be rigorously tested for robustness against failures.  

Radio Resource Control (RRC) is one of the most important protocols in telecom networks. It is a Layer 3 protocol and is used for radio resource management between mobile phones and a base station, which is the entry point to the network. Without graceful handling of RRC messages (either expected or unexpected), there would not be a communication channel between the mobile phones and the network. The RRC protocol also tunnels an upper layer protocol called Non-Access Stratum (NAS), meaning that the RRC messages impact mobile phones' communication with not only the base station but also other parts of the network beyond the base station. 

Consequences of problems in handling RRC and the tunneled NAS protocols are severe for both the networks as well as the mobile phones. Fixing mobile phones would require mass rollout of baseband over-the-air (OTA) updates which may not be possible for all types of mobile phones. Though relatively easier to fix on the network side (due to tighter control by network operators), the impact is much more severe because many users (hundreds or thousands) could lose service simultaneously until the problem is fixed. Therefore, it is imperative that the RRC protocol (including the tunneled NAS protocol) is handled in a robust manner by both the mobile phones and the network. 

First step in achieving this robustness is proper standardization in 3rd Generation Partnership Project (3GPP) \cite{3gpp}. But, when 3GPP technical specifications (TS) are implemented by different manufacturers of mobile phones and network equipments, factors like human errors or poor coding practices may introduce implementation bugs that go unnoticed. To uncover implementation bugs, these manufacturers need to perform two broad categories of software testing, i.e., conformance and fuzz testing. While conformance testing ensures that a System Under Test (SUT) works properly with expected messages, fuzz testing tries to identify vulnerabilities in the SUT by sending \underline{un}expected messages \cite{Takanen2009FuzzingT, Li2018FuzzingAS}. Fuzz testing complements conformance testing by discovering faults not identified by the latter, e.g., buffer overflow and race conditions. Discovery of faults contributes to a robust software by giving the manufacturers an opportunity to fix them before the software is shipped. Hence, fuzz testing of the RRC and the tunneled NAS protocols is indispensable to ensure robust telecom networks. 

\textbf{Related work.} Fuzz testing, in general, is a well-researched area. Some fuzzers support generic message format syntaxes, e.g., SPIKE \cite{immunitysecSpike}, Sulley \cite{openrceSulley}, Boofuzz \cite{jtpereydaBoofuzz}, SPFuzz \cite{Song2019SPFuzzAH}, SNOOZE \cite{banks2006snooze}, and Aspfuzz \cite{aspfuzz}. Others like AutoFuzz \cite{Gorbunov2010AutoFuzzAN} extract message format from recorded communication. These fuzzers -- while suitable for common IP-based protocols such as HTTP, SIP, and FTP -- are not suitable for the RRC protocol. The RRC messages are not only binary encoded but also tunneled in lower layer protocols. The lower layer protocols perform operations like header compression, encryption, integrity protection, and segmentation. So, fuzzing these messages carrying RRC messages with generic fuzzers will almost always produce garbage that is either invalid at the Layer 3 (RRC protocol layer) or already discarded by lower layers even before reaching the Layer 3. 

In context of general security protocols, to overcome the hurdle of lower layer protocol handlings like encryption, SecFuzz \cite{secfuzz} proposes to use a concrete implementation of an end-point towards a SUT and provide the fuzzer with necessary keys and algorithms. SecFuzz only covers the IKE protocol and is not directly applicable for the RRC protocol.

Telecom fuzzers like GSMFuzz \cite{Broek2014SecurityTO}, HFuzz \cite{hfuzz}, and the ones proposed by Cui et al. \cite{cui2015detection}, Yang et al. \cite{yang2016detection}, and Zhao et al. \cite{zhao2018fuzzing} cover protocols such as RLC, DIAMETER, GTP, and SMS, but not RRC and NAS. Then, there are fuzzers specific to the RRC and NAS protocols, which are most relevant to this paper. One of them is BaseSAFE \cite{maier2020basesafe} which fuzzes a mobile phone baseband by collecting downlink messages; using AFL++ \cite{aflpp} on those messages to generate a fuzzing corpus; and replaying the corpus to the mobile phone baseband. Another is LEFT \cite{fang2018emulation} that also fuzzes mobile phones but by perturbing the order of messages in a modified concrete implementation of a base station. Other works such as LTEInspector \cite{hussain2018lteinspector}, LTEFuzz \cite{Kim2019TouchingTU}, and 5GReasoner \cite{hussain20195greasoner} generate and use test cases based on manual analysis of 3GPP TSes.

A major hurdle with fuzzing the RRC protocol using the above-mentioned telecom fuzzers is that they are tied to the implementation of a particular 3GPP TS version. They are based on implementing or collecting RRC messages that conform to a particular version of 3GPP TS. This makes them non-backward compatible with older SUTs and soon-to-be obsolete with newer SUTs, because 3GPP TSes are continuously evolving for new features and enhancements. Therefore, re-implementation or re-collection of RRC messages would be inevitable. There are fuzzers like T-Fuzz \cite{Johansson2014TFuzzMF} for NAS and T3FAH \cite{Xu2009T3FAHAT} for SIP that are based on TTCN-3 schema and do not natively embed message formats. But they too are not sustainable for fuzzing the RRC protocol since RRC messages are not defined in 3GPP TSes using TTCN-3 format. BASESPEC \cite{Kim2021BASESPECCA} proposes Layer 3 testing by comparing baseband firmware to cellular specification, however, such comparison is mainly limited to conformance testing and not to fuzzing.

\textbf{Our work.} Despite its importance -- to our knowledge -- we see a research gap in fuzz testing of the RRC protocol that can cope with continuously evolving 3GPP TSes. To fill this research gap, we present the design and evaluation of our backward and forward compatible RRC fuzzer for 4G and 5G -- called \textit{Berserker}\footnote{Berserkers are historic Nordic warriors known to be extremely furious; the name represents our fuzzer that is ruthless with a system under test.}. 

Berserker relies on a concrete implementation of telecom protocol stack (\S \ref{berserker}). When fuzz testing the network side as SUT, it uses a concrete implementation of a mobile phone. Correspondingly, when fuzz testing the mobile phone side as SUT, it uses a concrete implementation of a network. In either case, the protocol handlings of the SUT are untouched. Berserker fuzzes RRC messages (including tunneled NAS) between the mobile phone and the network as below: acts as an additional shim layer in the concrete implementation between the Layer 3 and the lower layers; intercepts RRC messages in Layer 3 before they are passed down to lower layers; fuzzes or replaces the intercepted RRC messages; and passes down the fuzzed or replaced messages to lower layers. Thus, Berserker is unaffected by lower layer protocol handlings like encryption, integrity protection, segmentation, and scheduling. We note that besides being able to directly fuzz RRC messages, Berserker also provides a machinery to fuzz NAS protocol tunneled in RRC.

RRC message formats are defined in 3GPP TS 36.331 \cite{3gpp_36331} for 4G and 38.331 \cite{3gpp_38331} for 5G using Abstract Syntax Notation One (ASN.1), which is a platform agnostic language for defining data structures \cite{itu_X.680,itu_X.681}. In order to be backward and forward compatible with different versions of these TSes -- there are at least 175 versions of TS 36.331 from Release 8 to 16 and 20 versions of TS 38.331 from Release 15 to 16 -- we designed Berserker to obtain knowledge of RRC messages from whichever version it is provided with. This design makes Berserker stand out from existing designs that have message formats embedded into the fuzzer or need fuzzer specific formats or rely on concrete corpus. Berserker extracts RRC ASN.1 schema definitions directly from a 3GPP TS and compiles the extracted RRC ASN.1 schema to produce corresponding encoder and decoder.

However, using encoder and decoder based on the RRC ASN.1 schema would introduce one inconvenience; the encoder and decoder would follow constraints imposed by the ASN.1 schema on what values a field can take; e.g., if a field is defined to be 40-bits long, then the encoder would refuse to set that field to 80-bits long value. To overcome this inconvenience, we added a capability in Berserker to mutate the RRC ASN.1 schema itself, e.g., changing the field definition from 40-bits to 80-bits. The encoder produced from such mutated RRC ASN.1 schema would encode 80-bits fields, beyond the constrains imposed by original schema. Thus, Berserker can test the mobile phones and the network more rigorously.  

We evaluated Berserker with the open-source project, srsLTE \cite{srsLTE}, in two configurations (\S \ref{experiments}): (a) srsUE as the concrete implementation of 4G mobile phone; srsENB and srsEPC as the 4G network SUT, and (b) srsENB and srsEPC as the concrete implementation of 4G network; srsUE as the 4G mobile phone SUT. Berserker discovered two serious vulnerabilities in the srsEPC, thus proving its effectiveness  (\S \ref{findings_limitations}). One vulnerability lies in improper parsing and another in improper security processing of NAS messages that are tunneled in a RRC message. Upon further investigation, we learned that the improper parsing vulnerability is inherited from openLTE \cite{openlte_2019}, which is another open-source project from which srsLTE borrows some code.

To summarize, our contributions are:
\begin{enumerate}
	\item We designed a novel RRC fuzzer, \textit{Berserker}, that is both backward and forward compatible with continuously evolving 3GPP TSes by extracting RRC ASN.1 schema directly from any 3GPP RRC TS for 4G and 5G.
	\item We present a technique of sidestepping constraints in RRC ASN.1 schema by mutating the schema itself.	
	\item We evaluate the fuzzer using open-source components in srsLTE and discovered two previously unknown serious vulnerabilities in srsLTE, one of which also affects openLTE.
	\item We prove that RRC fuzzer can act a machinery to effectively fuzz upper layer NAS protocol.
	\item By using concrete implementations of telecom protocol stack, we show that the RRC (and NAS) layer can be fuzzed while staying unaffected by lower layer protocol handlings.
\end{enumerate}

\textbf{Responsible disclosure.} We have informed the srsLTE team about the crash. For wider disclosure to commercial manufacturers, other relevant open-source projects (including openLTE) as well as their users, we are coordinating with Ericsson’s PSIRT.

\section{RRC Protocol in 4G and 5G} \label{4g_5g_proto}
\begin{figure}[b]
	\centering
	\includegraphics[height=5cm]{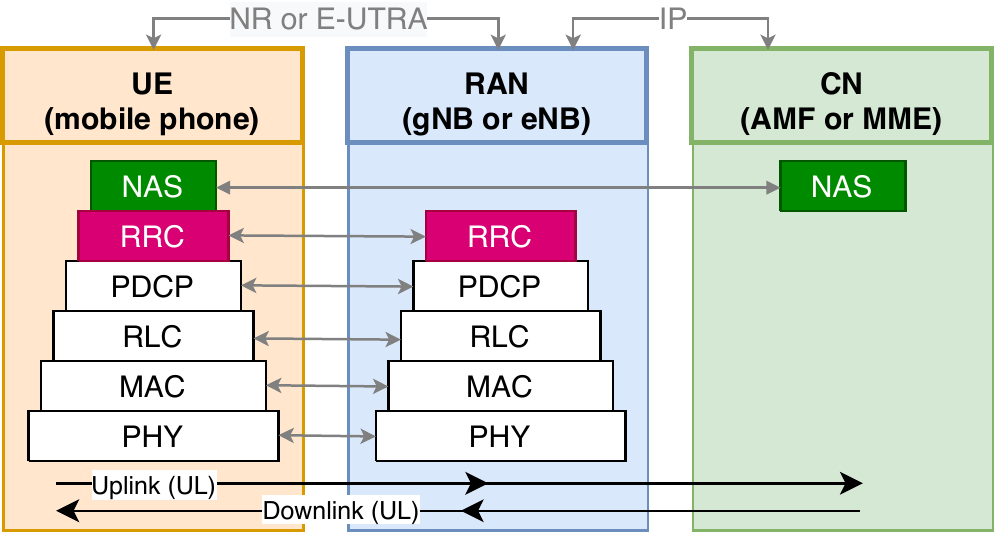}
	\caption {Control plane protocol stack used over the air interface in 4G and 5G.} 
	\label{fig:4G_5G_control_plane_stack} 
\end{figure}

4G and 5G are the fourth and fifth generations of telecom networks. Their architectures and protocols -- including security -- are standardized by 3GPP. On a high-level, they consist of three main entities as shown in Figure \ref{fig:4G_5G_control_plane_stack}: User Equipment (UE), Radio Access Network (RAN), and Core Network (CN). The UE represents mobile phones; the RAN provides wireless communication service to the UEs; the CN (among other functionalities) routes data between UEs and the Internet. The RAN consists of radio base stations that are called gNB in 5G and eNB in 4G. The radio access technology used by a gNB is called New Radio (NR), and that by an eNB is called Evolved Universal Terrestrial Radio Access (E-UTRA). The CN consists of several network functions among which Access and Mobility Management Function (AMF) in 5G and Mobility Management Entity (MME) in 4G are shown in the figure. Direction of messages from UE to RAN/CN is called uplink and the other direction is called downlink.

Traffic in 4G and 5G are of two types. One type -- called \textit{control plane} -- is responsible for establishing and maintaining connections, and setting up security. Another type -- called \textit{user plane} -- carries the actual data like voice calls and Internet data. Each type comprises of several protocols. In what follows, we briefly discuss the control plane protocol stack used over the air interface, comprising the Radio Resource Control (RRC) and Non-Access Stratum (NAS) protocols that this paper focuses on. 

As shown in Figure \ref{fig:4G_5G_control_plane_stack}, the Physical (PHY) layer, which is Layer 1, carries traffic using radio resources in time and frequency domains. Layer 2 consists of Medium Access Control (MAC), Radio Link Control (RLC), Packet Data Convergence Protocol (PDCP) layers. They enable functionalities such as uplink and downlink scheduling, packet segmentation, encryption, and integrity protection.

The RRC is a Layer 3 protocol that is responsible for creating and maintaining radio connections, broadcasting system information, delivering paging notifications, configuring and reporting radio measurements, and tunneling NAS messages. NAS is a control plane protocol between the UE and the CN; it is responsible for setting up security, and handling mobility and session management. 

In terms of functionalities, procedures, and interactions with upper and lower layers, the RRC protocol has largely remained the same in 4G (TS 36.331 \cite{3gpp_36331}) and 5G (TS 38.331 \cite{3gpp_38331}). However, the definitions of RRC messages -- including fields, types, and constraints -- have been continously evolving not only between 4G and 5G but also between different releases within 4G/5G. As stated earlier, there are at least 175 versions of 4G RRC TS from Release 8 to 16 and 20 versions of 5G RRC TS from Release 15 to 16.

\begin{figure}[h]
	\centering
	\includegraphics[width=10cm]{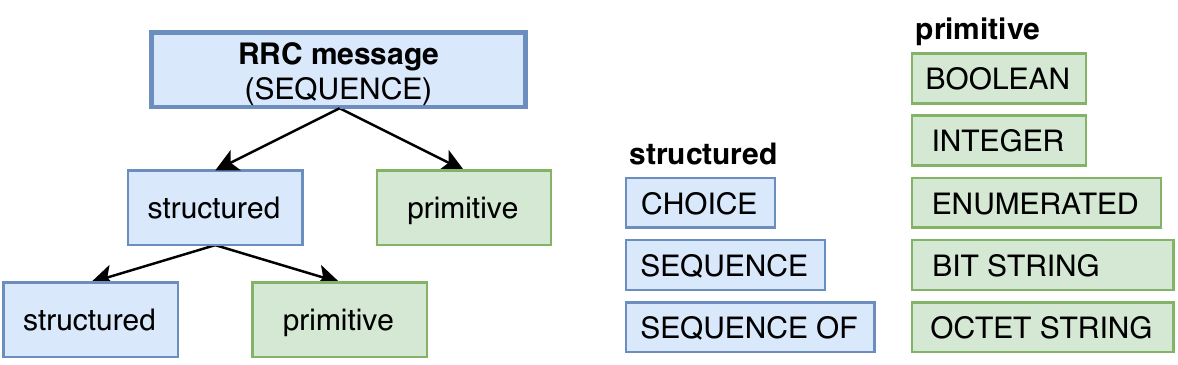}
	\caption{Illustration of RRC message defined using ASN.1.}
	\label{fig:sample_rrc_tree} 
\end{figure}

On a high-level, RRC messages are defined using platform agnostic ASN.1 and can be seen as a tree in Figure \ref{fig:sample_rrc_tree}. Each RRC message comprises of structured (SEQUENCE, SEQUENCE OF, CHOICE) and primitive (BOOLEAN, INTEGER, ENUMERATED, BIT STRING and OCTET STRING) data types. Each structured data type can further comprise of structured and primitive data types. During transfer, the RRC messages are encoded using Unaligned Packed Encoding Rules (UPER) \cite{itu_X.691} that converts data structures into compact series of bytes.

\section{Berserker: RRC fuzzer for 4G and 5G} \label{berserker}
When fuzzing the messages from UE to RAN/CN -- known as \textbf{uplink} messages -- the SUTs comprise of RAN and CN (Figure \ref{fig:4G_5G_control_plane_stack}); RAN is the SUT terminating RRC protocol; CN is the SUT terminating NAS protocol (originally tunneled in RRC). This uplink fuzzing is mainly useful for manufacturers of network equipments. In this case, Berserker relies on a concrete UE implementation while the RAN/CN’s protocol handlings are untouched.

Correspondingly, when fuzzing the messages from RAN/CN to UE -- known as \textbf{downlink} messages -- the SUT comprises of UE (Figure \ref{fig:4G_5G_control_plane_stack}); UE terminates both the RRC and NAS protocols. This downlink fuzzing is mainly useful for manufacturers of mobile phones. In this case, Berserker relies on a concrete RAN/CN implementation while the UE’s protocol handlings are untouched.

\begin{figure}[t]
	\centering
	\includegraphics[height=10cm]{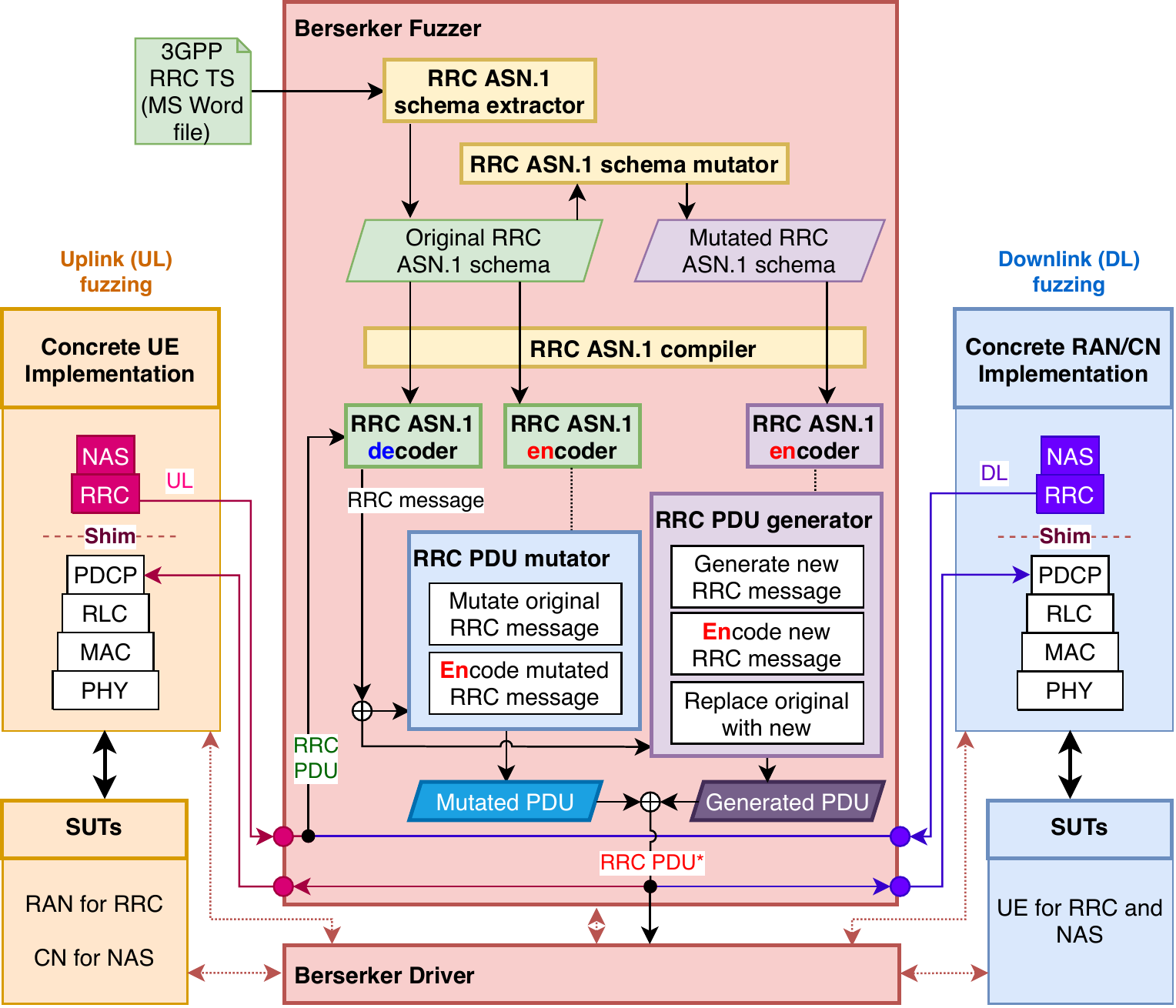}
	\caption{Components of Berserker. When RAN/CN are the SUTs, only the UE's UL messages are fuzzed while the RAN/CN's UL/DL protocol handlings are untouched. When UE is the SUT, only the RAN/CN's DL messages are fuzzed and the UE's UL/DL protocol handlings are untouched.}
	\label{fig:berserker} 
\end{figure}

Berserker comprises of two main components called \textit{Fuzzer} and \textit{Driver} as shown in Figure \ref{fig:berserker}. In what follows, we give detail description of these components.

\begin{figure}[h!]
	\centering
	\includegraphics[width=9cm]{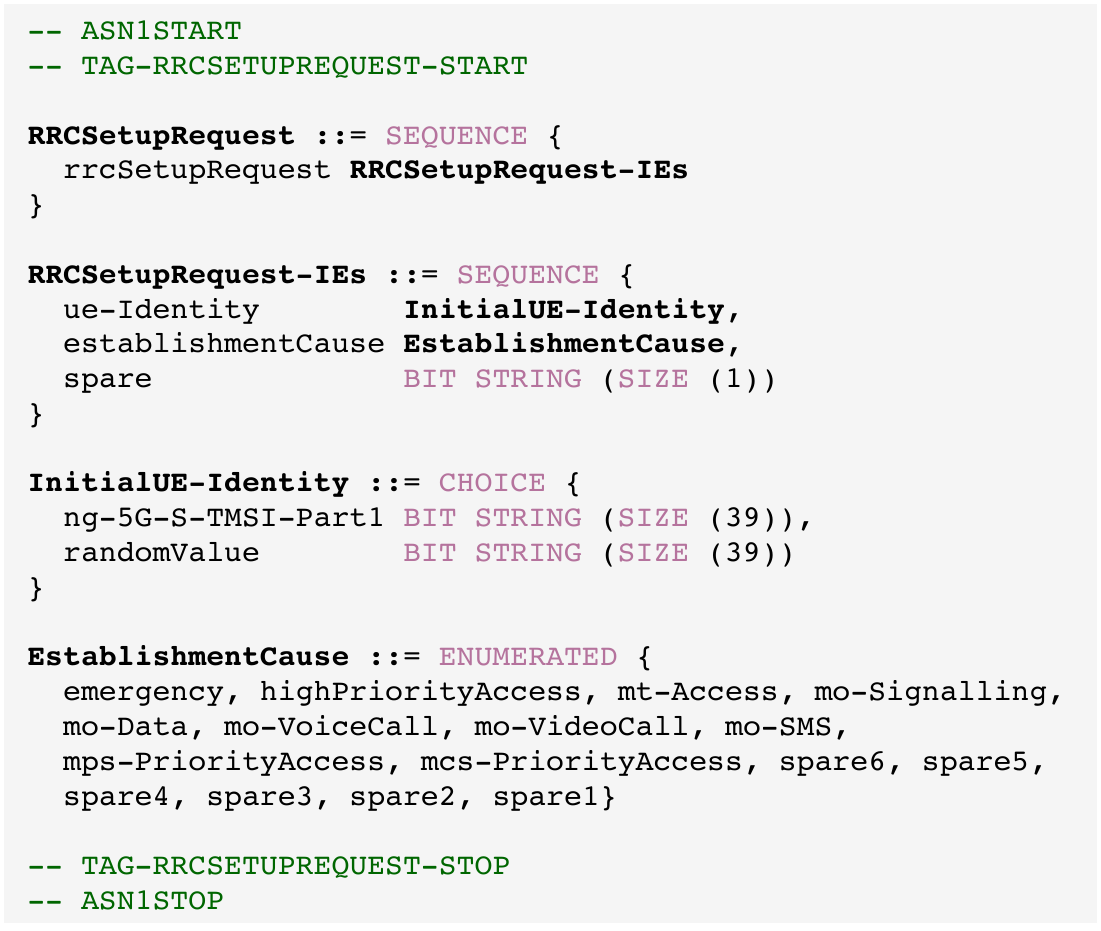}
	\caption{Example of a RRC message format defined using ASN.1 (snippet from 3GPP 38.331 \cite{3gpp_38331}).}
	\label{fig:rrc_con_req} 
\end{figure}

\subsection{Berserker Fuzzer}
Berserker Fuzzer component is responsible for fuzzing the RRC messages (including the tunneled NAS messages). It supports both: mutating the contents of an existing RRC message -- called \textit{mutation-based} fuzzing -- and generating a new RRC message from scratch -- called \textit{generation-based} fuzzing \cite{fsecure_fuzzing}. When doing mutation-based fuzzing, Berserker conforms to the ASN.1 constraints of the original RRC message. When doing generation-based fuzzing, Berserker replaces the original RRC message with a random RRC message which could be of a different type than the original message as well as not conforming to the original ASN.1 constraints.  

This component implements no part of RRC procedure handling and state management, instead relies on a concrete implementation of UE (for uplink fuzzing) and RAN/CN (for downlink fuzzing) to setup the communication in a state that RRC messages are sent to the SUT. It acts as an additional \textbf{shim} layer in the concrete implementations and mutates or replaces RRC messages before they are sent to lower layers and ultimately to the SUT. It comprises of following sub-components.

\textbf{RRC ASN.1 schema extractor.} This component extracts all RRC message definitions from a 3GPP RRC TS and produces an \textit{original RRC ASN.1 schema}. First, an appropriate version of RRC TS is manually identified based on what is suitable for testing, e.g., which version the concrete implementation of UE or RAN/CN, and the SUTs support. Different versions of RRC TSes for 4G and 5G are publicly available as MS Word files \cite{3gpp_36331, 3gpp_38331}. Next, this component extracts the complete RRC ASN.1 schema into a single plain text file which basically contains all the text paragraphs between an \texttt{ASN1TSTART} tag and the following \texttt{ASN1STOP} tag in the order they appear throughout the TS. Figure \ref{fig:rrc_con_req} shows an example of definition for \textit{RRC Setup Request} message. The \texttt{TAG-\textit{NAME}-START} and \texttt{TAG-\textit{NAME}-STOP} tags can be ignored.

\textbf{RRC ASN.1 schema mutator.} It mutates a RRC ASN.1 schema to sidestep the constraints in that schema. The intuition behind sidestepping the constraints in the schema is to increase chances of finding bugs in the SUTs by sending partially or completely non-conforming RRC messages. The SUTs will either handle such messages gracefully or a bug may be discovered; either way, the SUTs are tested more rigorously. This component takes the original RRC ASN.1 schema produced by the RRC ASN.1 schema extractor, performs various mutation strategies, and produces a \textit{mutated RRC ASN.1 schema}.  Mutation strategies for the RRC ASN.1 schema are listed in Table \ref{tab:mutation_strategies_asn1}. 

\textbf{RRC ASN.1 compiler.} This component takes the original and mutated RRC ASN.1 schema files, which are platform agnostic plain text files, and compiles them to produce programing language specific encoder and decoder of RRC messages. 

\textbf{RRC ASN.1 encoder and decoder.} Encoder converts programming language specific RRC message structures into ASN.1 UPER encoded series of bytes, and decoder does the opposite. They conform to constraints in the schema from which they were compiled. Berserker has one encoder and decoder pair compiled from the \textit{original} RRC ASN.1 schema, and another (only) encoder compiled from the \textit{mutated} RRC ASN.1 schema. So, there is one decoder that takes the intercepted RRC Protocol Data Unit (PDU) -- ASN.1 UPER encoded series of bytes -- from the concrete UE or RAN/CN implementation, and produces decoded RRC message. It is required for the decoder to conform to the original RRC ASN.1 schema because otherwise decoding of the intercepted RRC PDU will fail. There are two encoders. We discuss their use below.

\begin{table}[]
	\footnotesize
	\centering
	\caption{Mutation strategies for RRC ASN.1 schema}
	\label{tab:mutation_strategies_asn1}
	\renewcommand{\tabularxcolumn}[1]{m{#1}}
	\newcolumntype{L}{>{\raggedright\hsize=0.4\hsize\linewidth=\hsize}X}
	\newcolumntype{T}{>{\raggedright\arraybackslash\hsize=1.6\hsize\linewidth=\hsize}X}

	\begin{tabularx}{\linewidth}{|L|T|}
	\hline

	\textbf{Strategy} 
	& \textbf{Description} \textcolor{armygreen}{\textit{(examples relate to Figure \ref{fig:rrc_con_req})}}     
	\\ \hline

	Change primitive data types
	&Change type of data that are BOOLEAN, INTEGER, ENUMERATED, OCTET STRING, and BIT STRING.
	\textcolor{armygreen}{\textit{E.g., change spare from BIT STRING to BOOLEAN.}}
	\\ \hline
	
	Change structured data types
	&Change type of data that are SEQUENCE, SEQUENCE OF and CHOICE.
	\textcolor{armygreen}{\textit{E.g., change rrcSetupRequest from RRCSetupRequest-IEs to InitialUE-Identity.}}
	\\ \hline

	Extend options
	&Add new options for data that are ENUMERATED, SEQUENCE and CHOICE.
	\textcolor{armygreen}{\textit{E.g., add fuzz-code1 to EstablishmentCause.}}
	\\ \hline

	Reduce options
	&Remove existing options for data that are ENUMERATED, SEQUENCE and CHOICE.
	\textcolor{armygreen}{\textit{E.g., remove randomValue from InitialUE-Identity.}}	 
	\\ \hline

	Scramble options
	&Change the order of options for data that are SEQUENCE and CHOICE.
	\textcolor{armygreen}{\textit{E.g., change the order in RRCSetupRequest-IEs to spare, ue-Identity, establishmentCause.}}	 	     
	\\ \hline

	Change size
	&Change upper and lower limits of size for data that are INTEGER, OCTET STRING, BIT STRING, and SEQUENCE OF.
	\textcolor{armygreen}{\textit{E.g., change the size of ng-5G-S-TMSI-Part1 from 39 bits to 99 bits.}}
	\\ \hline

	\end{tabularx}
\end{table}

\begin{table}[t!]
	\footnotesize
	\centering
	\caption{Mutation and generation strategies for RRC messages}
	\label{tab:mutation_strategies_rrc}
	\renewcommand{\tabularxcolumn}[1]{m{#1}}
	\newcolumntype{L}{>{\raggedright\hsize=0.5\hsize\linewidth=\hsize}X}
	\newcolumntype{T}{>{\raggedright\arraybackslash\hsize=1.5\hsize\linewidth=\hsize}X}
	\newcolumntype{A}{>{\raggedright\hsize=2\hsize\linewidth=\hsize}X}

	\begin{tabularx}{\linewidth}{|L|T|}
	\hline
	
	\textbf{Data type}{\textsuperscript{a}}			
	& \textbf{Strategy}{\textsuperscript{b}} 	
	\\ \hline
	
	BOOLEAN \textcolor{armygreen}{\textit{(primitive)}}
	&Randomly set to True or False.
	\\ \hline

	INTEGER \textcolor{armygreen}{\textit{(primitive)}}                             
	&Set to a random integer within the corresponding SIZE constraints.
	\\ \hline

	ENUMERATED \textcolor{armygreen}{\textit{(primitive)}} 
	&Set to a random chosen option from corresponding enum options.
	\\ \hline

	BIT STRING\newline\textcolor{armygreen}{\textit{(primitive)}}	
	&If bounded by SIZE constraints, set to a random bit series of random length within the SIZE constraints. Otherwise, if unbounded, set to a random bit series of length 8000{\textsuperscript{c}}. (or use radamsa{\textsuperscript{d}})		
	\\ \hline

	OCTET STRING\newline\textcolor{armygreen}{\textit{(primitive)}}	
	&If bounded, set to a random octet series of random length within the SIZE constraints. Otherwise, if unbounded, set to a random octet series of length 1000{\textsuperscript{c}}. (or use radamsa{\textsuperscript{d}})
	\\ \hline

	SEQUENCE\newline\textcolor{armygreen}{\textit{(structured)}}
	&For mutation, traverse to find primitive data types, and apply corresponding strategy above when found. For generation, traverse for structured and primitive data types, and initialize them with corresponding strategy above.
	\\ \hline

	SEQUENCE OF\newline\textcolor{armygreen}{\textit{(structured)}}
	&For mutation, traverse to find primitive data types, and apply corresponding strategy above when found. For generation, initialize a random number of corresponding structured or primitive data type with corresponding strategy above.	
	\\ \hline

	CHOICE\newline\textcolor{armygreen}{\textit{(structured)}}
	&For mutation, traverse to find primitive data types, and apply corresponding strategy above when found. For generation, randomly choose one of corresponding options and initialize the chosen structured or primitive data type with corresponding strategy above.	
	\\ \hline

	OPTIONAL\newline\textcolor{armygreen}{\textit{(qualifier)}}
	&For mutation, remove included field or include missing field. For generation, randomly choose to include field. If included, initialize with corresponding strategy{\textsuperscript{e}} above.
	\\ \hline

	Specific name\newline\textcolor{armygreen}{\textit{(custom, for all data types)}}
	&In mutation, a specific field name can be chosen if multiple fields are of same type. Not used in generation.
	\\ \hline

	Perturbation\newline\textcolor{armygreen}{\textit{(custom, for messages)}}
	&Not used in mutation. In generation, messages are replaced out-of-order. 
	\\ \hline
	
	\multicolumn{2}{|l|}{\makecell[cl]{\textsuperscript{a}Working on ASN.1 data types instead of concrete RRC fields makes Berserker generic to any version of 4G and 5G 3GPP \\\hspace{0.1cm}RRC TSes.}}
	\\ 
	\multicolumn{2}{|l|}{\makecell[cl]{\textsuperscript{b}Strategies for both the mutation- and generation-based fuzzing always conform to the constraints of underlying ASN.1 \\\hspace{0.1cm}schema. In generation-based fuzzing, sidestepping the constraints comes from the fact that the underlying schema is \\\hspace{0.1cm}mutated, not from non-conforming strategy.}}
	\\ 	
	\multicolumn{2}{|l|}{\makecell[cl]{\textsuperscript{c}The chosen length is based on our empirical observation with srsLTE that messages above certain length do not get \\\hspace{0.1cm}transferred, although up to 8188 octets should have been supported \cite{3gpp_eutra_pdcp}.}}
	\\ 	
	\multicolumn{2}{|l|}{\makecell[cl]{\textsuperscript{d}radamsa is an open-source general-purpose fuzzer \cite{radamsa}.}}	
	\\
	\multicolumn{2}{|l|}{\makecell[cl]{\textsuperscript{e}In a so-called Christmas tree message, \textbf{all} optional fields are included.}}	
	\\ \hline
	\end{tabularx}
\end{table}

\textbf{RRC PDU mutator.} It is responsible for mutation-based fuzzing in which the contents of an \textit{existing} message are modified. It takes the decoded RRC message from the decoder; performs various mutation strategies listed in Table \ref{tab:mutation_strategies_rrc}; and produces a mutated PDU (called RRC PDU\textbf{*}) using the RRC ASN.1 encoder compiled from the original RRC ASN.1 schema. This RRC PDU* is then submitted to lower layers.

We note that the encoder used by this mutator component conforms to the original RRC ASN.1 schema. Unlike the decoder, although conforming to the original RRC ASN.1 schema is not strictly required for the encoder, it is our design choice to focus the mutation-based fuzzing within the same constraints as the intercepted RRC PDU. This way, it is guaranteed that the RRC PDU* will never be early rejected by the SUTs (assuming robust SUTs) because of non-conformance. 

\textbf{RRC PDU generator.} This component does generation-based fuzzing in which a \textit{new} message is generated from scratch. It instantiates a randomly chosen empty RRC message; fills the content based on various generation strategies listed in Table \ref{tab:mutation_strategies_rrc}; generates a RRC PDU\textbf{*} using the RRC ASN.1 encoder compiled from the mutated RRC ASN.1 schema; and replaces the original RRC message with the RRC PDU\textbf{*}. This RRC PDU* is submitted to lower layers.

This component performs two types of tests on the SUTs. One type is to test how the SUTs handle RRC messages (including tunneled NAS) deviating from constraints in the original RRC ASN.1 schema. This enables testing of even the SUTs' implementation of ASN.1 decoder as well as NAS handler. In this case, the encoder used by this component conforms to the mutated RRC ASN.1 schema. 

Another type is to test how the SUTs handle out-of-order messages (also known as perturbation, order-level message fuzzing, or sequence fuzzing). In this case, we made a design choice to \textit{not} mutate the schema, meaning that the encoder is as good as conforming to the original RRC ASN.1 schema. By doing so, the RRC PDU* will not be early rejected by the SUT on account of non-conformance. 

\subsection{Berserker Driver}
Berserker Driver component is responsible for monitoring the concrete UE or RAN/CN implementations, and the SUTs; and providing feedback to the Berserker Fuzzer component. It also orchestrates the fuzzing by e.g., controlling seed for randomness; configuring mutation strategies for ASN.1 schema; configuring mutation and generation strategies for RRC messages; triggering transfer of RRC messages; restarting the concrete UE or RAN/CN implementations, and the SUTs if unresponsive.
\section{Experiments} \label{experiments}
\textbf{Components.} In order to evaluate Berserker's design (Figure \ref{berserker}), we did experiments by using the components listed in Table \ref{tab:experiment_env}. Among them, asn1c \cite{velichkov} and srsLTE \cite{srsLTE} are the most prominent. asn1c is an open-source ASN.1 compiler that converts ASN.1 schema into C source code (compatible with C++). srsLTE is an open-source implementation of 4G protocol stack written in C++ and comprising of a CN (called srsEPC), a RAN (called srsENB), and a UE (called srsUE). We chose srsLTE because not only it is popular among telecom security researchers, but also an open-source implementation of 5G protocol stack was unavailable at the time of writing (OAI's 5G RAN \cite{oai_ran} and 5G CN \cite{oai_cn} are not yet 5G ready). We note that although our experiments are in 4G, they are sufficiently general because the control plane protocol stack (Figure \ref{4g_5g_proto}) are in same order between 4G and 5G, and the RRC protocol in both are defined using ASN.1. For mutation-based fuzzing of BIT STRING and OCTET STRING, we used radamsa \cite{radamsa} in addition to random data. 

\begin{table}[ht]
	\footnotesize
	\centering
	\caption{Experiment components}
	\label{tab:experiment_env}
	\renewcommand{\tabularxcolumn}[1]{m{#1}}
	\newcolumntype{L}{>{\raggedright\hsize=0.45\hsize\linewidth=\hsize}X}
	\newcolumntype{T}{>{\raggedright\arraybackslash\hsize=1.55\hsize\linewidth=\hsize}X}
	\newcolumntype{A}{>{\raggedright\hsize=2\hsize\linewidth=\hsize}X}

	\begin{tabularx}{\linewidth}{|L|T|}
	\hline

	\textbf{Component}
	& \textbf{Details}
	\\ \hline

	SUTs
	& srsENB/srsEPC (for uplink fuzzing) and srsUE (for downlink fuzzing) applications in srsLTE{\textsuperscript{a}}. srsENB terminates RRC, srsEPC terminates NAS, and srsUE terminates both RRC and NAS. Inactivity timer at srsENB was set to 5 seconds.
	\\ \hline
	
	Concrete implementations
	& srsUE (for uplink fuzzing) and srsENB/srsEPC (for downlink fuzzing) applications in srsLTE{\textsuperscript{b}}. They generate RRC and NAS (tunneled in RRC) messages and send them to SUTs.
	\\ \hline	

	RF front-end
	& ZeroMQ-based RF driver{\textsuperscript{c}} provided by srsLTE. It exchanges IQ samples over TCP. 
	\\ \hline	

	3GPP RRC TS
	& 3GPP TS 36.331 \cite{3gpp_36331} version 15.4.0 (2019-02-19). It was chosen based on what srsLTE supports. It is a MS Word file sized 11.9 MB.
	\\ \hline

	RRC ASN.1 schema extractor
	& Our inhouse tool{\textsuperscript{d}}. It extracts RRC ASN.1 schema from a 3GPP RRC TS. The extracted schema from the above TS is a plain text file sized 668 KB.
	\\ \hline
	
	RRC specification mutator
	& A Python (version 3.8) script. It produces a plain text file with mutated RRC ASN.1 schema. There are multiple mutated files.
	\\ \hline

	RRC ASN.1 compiler
	& asn1c. We use a fork{\textsuperscript{e}} \cite{velichkov} of original asn1c to simultaneously work with encoders and decoders compiled from different ASN.1 schemas.
	\\ \hline

	RRC encoders and decoder
	& Namespace-separated custom wrapper around C files produced by the fork of asn1c.
	\\ \hline

	Shim, RRC mutator, and generator
	& Integrated into \texttt{send\_<ul,dl>\_ccch} and \texttt{send\_<ul,dl>\_dcch} functions in srsUE's \texttt{src/stack/rrc/rrc.cc} (for uplink fuzzing) and srsENB's \texttt{src/stack/rrc/rrc.cc} (for downlink fuzzing). radamsa{\textsuperscript{f}} is also integrated here.
	\\ \hline

	Driver
	& Combination of bash scripts and configuration files. 
	\\ \hline

	Test machines (virtual)
	& 4 x (4GB RAM, Intel Xeon E3-12xx v2 of x86\_64), ~~ 1 x (10GB RAM, Intel(R) Core(TM) i5-8350u CPU @ 1.70ghz) all using Ubuntu 18.04 64-bit with gcc version 7.5.0. Each virtual machine had a complete experiment setup.
	\\ \hline

	\multicolumn{2}{|l|}{\makecell[cl]{\textsuperscript{a, b}srsLTE -- c892ae56be5302eaee5ca00e270efc7a5ce6fbb2 commit tag, Release 20.04.1.}}
	\\
	\multicolumn{2}{|l|}{\makecell[cl]{\textsuperscript{c}An actual RF front-end and transmission on the air interface is unnecessary for our purpose of fuzzing the RRC protocol \\\hspace{0.1cm}since Berserker is unaffected by lower layer protocols.}}
	\\
	\multicolumn{2}{|l|}{\makecell[cl]{\textsuperscript{d}An alternative publicly available tool is at \cite{oai_asn}. Clause A.3.1.1 in \cite{3gpp_38331} describes how one can make their own extractor.}}
	\\
	\multicolumn{2}{|l|}{\makecell[cl]{\textsuperscript{e}asn1c fork -- 27eaf82abed937a2da5a5fd9e0d4076d9abcfc75 commit tag.}}
	\\
	\multicolumn{2}{|l|}{\makecell[cl]{\textsuperscript{f}radamsa -- d71c384bafb53865a561684035fbea1cf6c76734 commit tag.}}
	\\ \hline

	\end{tabularx}
\end{table}

\begin{figure}[ht!]
	\centering
	\includegraphics[width=10cm]{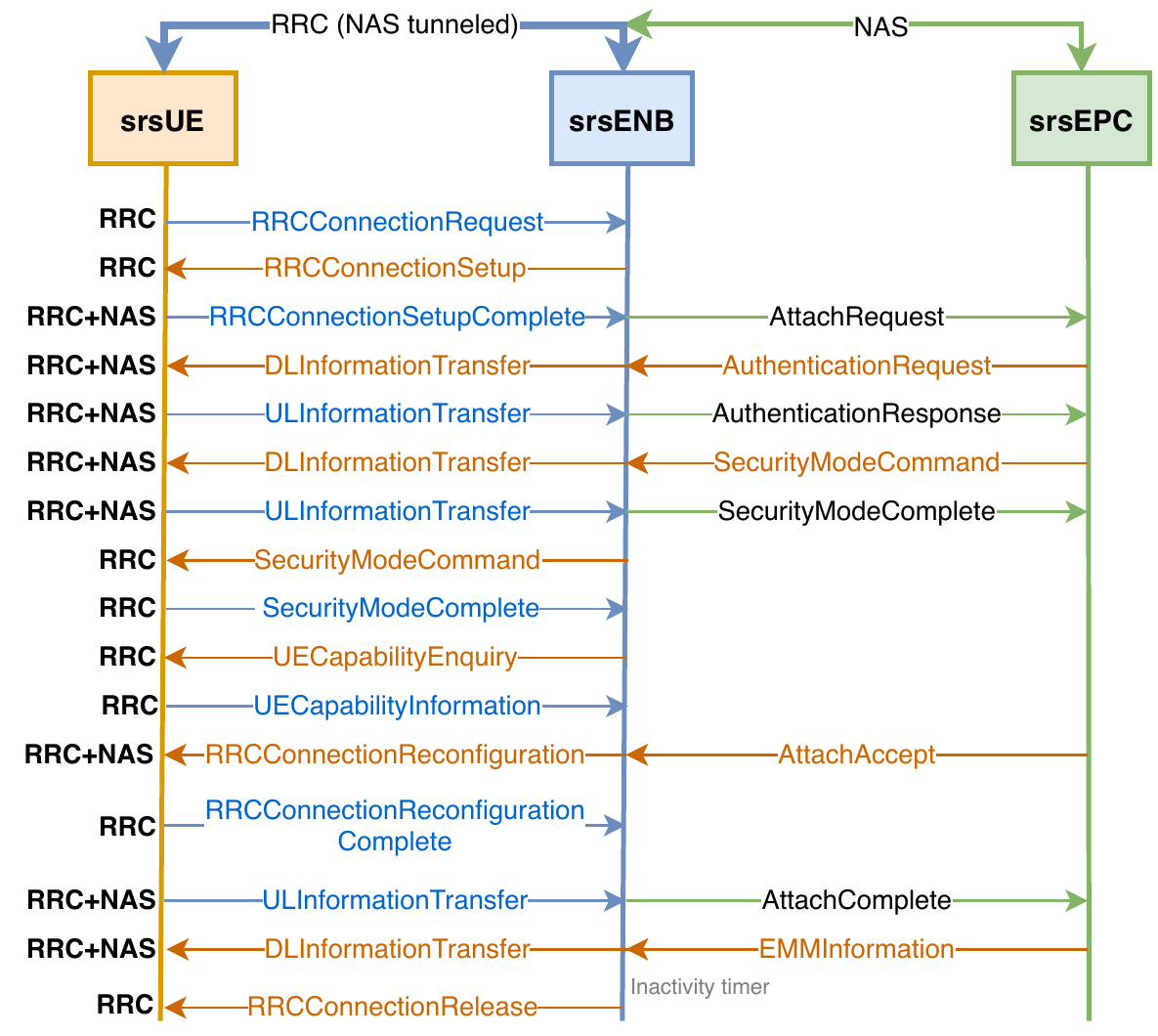}
	\caption {RRC and NAS messages (uplink and downlink) observed in our experiment with srsLTE.} 
	\label{fig:rrc_sequence} 
\end{figure}

\textbf{RRC and NAS messages in our experiments.} Figure \ref{fig:rrc_sequence} shows the uplink and downlink messages we observed during an initial attach procedure in srsLTE. In uplink, there are six RRC messages (RRCConnectionRequest, RRCConnectionSetupComplete, ULInformationTransfer, SecurityModeComplete, UECapabilityInformation, and RRCConnectionReconfigurationComplete) and four NAS messages (AttachRequest, AuthenticationResponse, SecurityModeComplete, and AttachComplete). In downlink too, there are six RRC messages (RRCConnectionSetup, DLInformationTransfer, SecurityModeCommand, UECapabilityEnquiry, RRCConnectionReconfiguration, and RRCConnectionRelease) and four NAS messages (AuthenticationRequest, SecurityModeCommand, AttachAccept, and EMMInformation). When doing mutation-based fuzzing, the RRC messages (including tunneled NAS messages) are modified. When doing generation-based fuzzing, they are replaced with any other RRC message (that may include a tunneled NAS message) available in the RRC ASN.1 schema.

\textbf{Collecting results.} The Driver continuously triggers RRC messages by either (re)starting the srsUE or by initiating an IP ping request via srsUE. For the purpose of analysis, the Driver stores all the mutated and generated RRC PDUs produced by the Fuzzer (PDU* in Figure \ref{berserker}) and application logs from srsUE, srsENB and srsEPC. The Driver restarts all the three applications (srsUE, srsENB and srsEPC) whenever any of them becomes unresponsive or crashes. It is so because sometimes, even when only one of the applications crashes, no messages could be communicated between srsUE and srsENB. The Driver tracks initialization seed for each experiment run so that tried values/paths are excluded when the applications are restarted.

Results from mutation-based fuzzing are given in Table \ref{tab:results_fuzzing_strategies}. Each row signifies the strategies used by Berserker's RRC PDU mutator. Similarly, results from generation-based fuzzing are given in Tables \ref{tab:results_generation_strategies_con_req} - \ref{tab:results_generation_strategies_rrc_conn_rel}. In each row, we identify the possible mutations on ASN.1 schema following Table \ref{tab:mutation_strategies_asn1}; during perturbation, in addition to mutating ASN.1 schemas, we sent other RRC messages that are generated using original ASN.1 schema.
\\

During mutation-based fuzzing,
\begin{itemize}[noitemsep,topsep=2pt,parsep=0pt,partopsep=0pt]
	\item 44,383 uplink RRC messages were mutated over a cumulative period of 13 days, 7 hours, and 41 minutes.
	\item 17,293 downlink RRC messages were mutated over a cumulative period of 11 days, 8 hours, and 22 minutes.
\end{itemize}

During generation-based fuzzing,
\begin{itemize}[noitemsep,topsep=2pt,parsep=0pt,partopsep=0pt]
	\item 2,598 uplink RRC messages were replaced over a cumulative period of 1 day, 3 hours, and 51 minutes.
	\item 1,997 downlink RRC messages were replaced over a cumulative period of 1 day, 2 hours, and 9 minutes.
\end{itemize}

We want to point that the total run times are \textit{not} indicative of one of mutation- and generation-based methods being faster or better than another; it was just our choice to run mutation-based fuzzing longer with higher stop value for seeds. The purpose of seeds shown in the results is twofold: first, they make the experiments reproducible by serving as initialization values for randomizations; and second, they limit the number of iterations. We manually choose stop value for seeds (also) depending on what is being fuzzed, e.g., only 8 iterations are sufficient for an ENUMERATED type with 8 options; and only few hundreds of values are iterated instead of all possible values (\(2^{40}\)) for a BIT STRING of SIZE 40. We also note that the number of fuzzed or replaced RRC messages vary even for the same stop seeds. It is because the srsUE sometimes performs several reconnection attempts itself without trigger from the Driver.

\begin{table*}[h!]
	\fontsize{6.5}{7} \selectfont
	\centering
	\caption{Results from mutation-based fuzzing -- both uplink and downlink messages}
	\label{tab:results_fuzzing_strategies}
	\renewcommand{\tabularxcolumn}[1]{m{#1}}
	\newcolumntype{A}{>{\raggedright\hsize=1.7\hsize\linewidth=\hsize}X}
	\newcolumntype{B}{>{\centering\arraybackslash\hsize=0.5\hsize\linewidth=\hsize}X}
	\newcolumntype{C}{>{\raggedright\hsize=2.75\hsize\linewidth=\hsize}X}
	\newcolumntype{D}{>{\centering\arraybackslash\hsize=0.4\hsize\linewidth=\hsize}X}
	\newcolumntype{E}{>{\centering\arraybackslash\hsize=0.65\hsize\linewidth=\hsize}X}
	\newcolumntype{F}{>{\centering\arraybackslash\hsize=0.7\hsize\linewidth=\hsize}X}
	\newcolumntype{G}{>{\centering\arraybackslash\hsize=0.3\hsize}X}

	\begin{tabularx}{\textwidth}{|A|B|C|D|E|F|G|}
		
		\hline\textbf{RRC message} & \textbf{Direction} & \textbf{Fuzzing strategies} & \textbf{No. of RRC PDUs mutated} & \textbf{Time taken to test (days, hours, minutes)} & \textbf{Added time for fuzzing (milliseconds)} & \textbf{Seeds used} \\ \hline
		
		\multirow{7}{*}{RRCConnectionRequest} & \multirow{7}{*}{Uplink}       & randomValue (BIT STRING 40) & 507 & 9:26 & 0.35-4.33 & 0-100 \\ \cline{3-7}
		&& mmec (BIT STRING 8) & 7575 & 1 day 4:40 & 0.29-5.32 & 0-100 \\ \cline{3-7}
		&& m-TMSI (BIT STRING 32) & 5787 & 2 days 4:35 & 0.28-4.09 & 0-100 \\ \cline{3-7}
		&& ng-5G-S-TMSI-Part1 (BIT STRING 40) & \multicolumn{4}{l|}{Not available in the srsLTE evaluation setup}\\ \cline{3-7}
		&& ENUM & 112 & 00:26 & 0.29-2.68 & 0-10 \\ \cline{3-7}
		&& BIT STRING & 4787 & 1 day 21:53 & 0.35-4.03 & 0-100 \\ \cline{3-7}
		&& radamsa fuzzing BIT STRINGS  & 99 & 1:24 & 4.00-166.84 & 0-100 \\ \hline

		\multirow{5}{*}{RRCConnectionSetup} & \multirow{5}{*}{Downlink} & INTEGER & 193 & 3:06 & 0.86-30.34 & 0-50 \\ \cline{3-7}
		&& Include optional fields & 50 & 4:34 & 27.29-102.16 & 0-50 \\ \cline{3-7}
		&& ENUM & 120 & 1:44 & 0.93-7.34 & 0-50 \\ \cline{3-7}
		&& Boolean & 954 & 03:39 & 0.79-119.61 & 0-50 \\ \cline{3-7}
		&& INTEGER, ENUM and Boolean & 79 & 1:26 & 1.08-23.60 & 0-50 \\ \hline
		
		\multirow{8}{*}{RRCConnectionSetupComplete} & \multirow{8}{*}{Uplink} & INTEGER & 110 & 00:23 & 0.29-3.17 & 0-10 \\ \cline{3-7}
		&& OCTET STRING & 107 & 2:34 & 0.21-2.05 & 0-100 \\ \cline{3-7}
		&& Append to dedicatedInfoNAS & 99 & 2:31 & 0.23-1.62 & 0-100 \\ \cline{3-7}
		&& Append to dedicatedInfoNAS and include optional fields & 100 & 2:31 & 0.51-12.49 & 0-100 \\ \cline{3-7}
		&& Include optional fields & 1740 & 1 day 8:48 & 0.64-29.60 & 0-100 \\ \cline{3-7}
		&& Fuzz existing and append to dedicatedInfoNAS, include optional fields & 98 & 2:41 & 0.57-3.30 &  0-100 \\ \cline{3-7}
		&& INTEGER and OCTET STRING & 50 & 1:16 & 0.23-2.47 & 0-50 \\ \cline{3-7}
		&& radamsa fuzzing OCTET STRING & 98  & 1:24 & 0.61-21.03 & 0-100 \\ \hline
		
		\multirow{8}{*}{DLInformationTransfer} & \multirow{8}{*}{Downlink} & INTEGER & 319 & 23:51 & 0.47-4.94 & 0-50 \\ \cline{3-7}
		&& Include optional fields & 303 & 2:43 & 0.51-6.23 & 0-50 \\ \cline{3-7}
		&& OCTET STRING & 299 & 6:12 & 0.67-9.65 & 0-50 \\ \cline{3-7}
		&& Append to dedicatedInfoNAS & 595 & 5:34 & 0.48-6.68 & 0-50 \\ \cline{3-7}
		&& Append to dedicatedInfoNAS and include optional fields & 604 & 5:37 & 0.51-8.23 & 0-50 \\ \cline{3-7}
		&& Append to and fuzz dedicatedInfoNAS; include optional fields & 304 & 6:14 & 0.72-6.63 & 0-50 \\ \cline{3-7}
		&& INTEGER and OCTET STRING & 300 & 6:14 & 0.67-8.34 & 0-50 \\ \cline{3-7}
		&& radamsa fuzzing OCTET STRING & 105 & 2:13 & 0.67-5.39   & 0-100 \\ \hline
		
		\multirow{6}{*}{ULInformationTransfer} & \multirow{6}{*}{Uplink} & OCTET STRING & 257 & 2:58 & 0.21-5.34 & 0-100 \\ \cline{3-7}
		&& Append to dedicatedInfoNAS & 437 &2:47 & 0.30-4.92 & 0-100 \\ \cline{3-7}
		&& Append to dedicatedInfoNAS and include optional fields & 164 & 5:47 &0.30-6.92 & 0-100 \\ \cline{3-7}
		&& Include optional fields & 156 & 6:38 &0.24-2.60 & 0-100 \\ \cline{3-7}
		&& Fuzz existing and append to dedicatedInfoNAS,  include optional fields & 256 & 3:07 & 0.34-12.78 & 0-100 \\ \cline{3-7}
		&& radamsa fuzzing OCTET STRING & 235 & 1:39 & 0.53-8.73 & 0-100 \\ \hline

		\multirow{4}{*}{SecurityModeCommand} & \multirow{4}{*}{Downlink} & INTEGER & 1270 & 20:28 & 5.47-379.60 & 0-50 \\ \cline{3-7}
		&& Include optional fields & 609 & 6:18 & 0.57-650.96 & 0-50 \\ \cline{3-7}
		&& ENUM & 54 & 3:04 & 0.60-4.16 & 0-50 \\ \cline{3-7}
		&& INTEGER and ENUM & 56 & 3:02 & 0.60-3.12 & 0-50 \\ \hline
		
		\multirow{2}{*}{SecurityModeComplete} & \multirow{2}{*}{Uplink} & INTEGER & 101 & 0:23 & 0.26-2.65 & 0-10 \\ \cline{3-7}
		&& Include Optional fields & 6049 & 1 day 1:30 & 0.29-2.06 & 0-100 \\ \hline

		\multirow{4}{*}{UECapabilityEnquiry} & \multirow{4}{*}{Downlink} & INTEGER & 1318 & 21:55 & 0.47-11.30 & 0-50 \\ \cline{3-7}
		&& Include optional fields & 102 & 3:52 & 2.07-338.23 & 0-50 \\ \cline{3-7}
		&& ENUM & 685 & 7:19 & 0.50-10.96 & 0-50 \\ \cline{3-7}
		&& INTEGER and ENUM & 701 & 7:10 & 0.52-4.36 & 0-50 \\ \hline
		
		\multirow{9}{*}{UECapabilityInformation} & \multirow{9}{*}{Uplink} & INTEGER & 67 & 0:19 & 0.29-2.08 & 0-10 \\ \cline{3-7}
		&& ENUM & 72 & 0:18 & 0.33-2.03 & 0-10 \\ \cline{3-7}
		&& OCTET STRING & 81 & 2:28 & 0.23-3.23 & 0-100 \\ \cline{3-7}
		&& Append to ueCapabilityRAT-Container & 6431 & 20:58 & 0.32-4.14 & 0-100 \\ \cline{3-7}
		&& Append to ueCapabilityRAT-Container and include optional fields & 2905 & 16:40 & 0.46-5.01 & 0-100 \\ \cline{3-7}
		&& Fuzz existing and append to ueCapabilityRAT- Container, include optional fields & 93 & 2:31 & 0.41-3.66 & 0-100 \\ \cline{3-7}
		&& Include optional fields & 3098 & 19:10 & 0.35-4.59 & 0-100 \\ \cline{3-7}
		&& INTEGER and ENUM & 1431 & 6:27 & 0.34-3.77 & 0-50 \\ \cline{3-7}
		&& radamsa fuzzing OCTET STRING & 133 & 2:12 & 0.56-5.71 & 0-100 \\ \hline
		
		\multirow{12}{*}[+1.5em]{RRCConnectionReconfiguration} & \multirow{12}{*}[+1.5em]{Downlink} & INTEGER & 1527 & 1 day 5:24 & 0.81-79.67 & 0-50 \\ \cline{3-7}
		&& OCTET STRING & 296 & 6:15 & 0.57-23.70 & 0-50 \\ \cline{3-7}
		&& INTEGER and OCTET STRING & 736 & 9:37 & 0.80-158.81 & 0-50 \\ \cline{3-7}
		&& ENUM & 663 & 8:04 & 0.85-36.47 & 0-50 \\ \cline{3-7}
		&& Boolean & 646 & 6:08 & 74-37.28 & 0-50 \\ \cline{3-7}
		&& INTEGER and ENUM & 776 & 6:08 & 0.74-37.28 & 0-50 \\ \cline{3-7}
		&& ENUM and OCTET STRING & 662 & 8:05 & 0.85-37.30 & 0-50 \\ \cline{3-7}
		&& INTEGER, ENUM and Boolean & 288 & 00:49 & 0.94-88.54 & 0-50 \\ \cline{3-7}
		&& radamsa fuzzing OCTET STRING & 100 & 1:03 & 0.77-29.82 & 0-100 \\ \hline


		\multirow{2}{*}{\makecell[cl]{RRCConnectionReconfiguration\\Complete}} & \multirow{2}{*}{Uplink} & INTEGER & 121 & 0:47 & 0.56-4.17 & 0-10 \\ \cline{3-7}
		&& Include Optional fields & 1592 & 20:55 & 0.67-41.14 & 0-100 
		\\ \hline

		\multirow{4}{*}{RRCConnectionRelease} & \multirow{4}{*}{Downlink} & INTEGER & 1374 & 1 day 1:34 & 0.47-53.11 & 0-50 \\ \cline{3-7}
		&& Include optional fields & 70 & 4:00 & 23.08-81.17 & 0-50 \\ \cline{3-7}
		&& ENUM & 640 & 6:37 & 0.54-16.53 & 0-50 \\ \cline{3-7}
		&& INTEGER and ENUM & 700 & 7:25 & 5.97-89.67 & 0-50 \\ \hline
	
\end{tabularx}
\end{table*}

\begin{table*}[ht!]
	\scriptsize
	\centering
	\caption{Results from generation-based fuzzing -- RRCConnectionRequest}
	\label{tab:results_generation_strategies_con_req}
	\renewcommand{\tabularxcolumn}[1]{m{#1}}
	\newcolumntype{A}{>{\raggedright\hsize=1.1\hsize\linewidth=\hsize}X}
	\newcolumntype{B}{>{\raggedright\hsize=3.25\hsize\linewidth=\hsize}X}
	\newcolumntype{C}{>{\centering\arraybackslash\hsize=0.45\hsize\linewidth=\hsize}X}
	\newcolumntype{D}{>{\centering\arraybackslash\hsize=0.45\hsize\linewidth=\hsize}X}
	\newcolumntype{E}{>{\centering\arraybackslash\hsize=0.5\hsize\linewidth=\hsize}X}
	\newcolumntype{F}{>{\centering\arraybackslash\hsize=0.25\hsize\linewidth=\hsize}X}

	\begin{tabularx}{\textwidth}{|A|B|C|D|E|F|}
		\hline\textbf{ASN.1 specification mutation} & \textbf{Mutation selected} & \textbf{Number of RRC PDUs replaced} & \textbf{Time taken to test (hours, minutes)} & \textbf{Added time for fuzzing (milliseconds)} &\textbf{Seeds used} \\ \hline

		\multirow{3}{*}{\makecell[cl]{Change primitive data \\type of randomValue \\(from BIT STRING)}} & to OCTET STRING & 12 & 00:15 & 1.7-2.9 & 0-10 \\ \cline{2-6}
		
		& to INTEGER & 12 & 00:15 & 1.5-6.1 & 0-10 \\ \cline{2-6}

		& to BOOLEAN & 13 & 00:15 & 0.7-3.6 & 0-10 \\ \hline

		Change structured types & from InitialUE-Identity to InitialUE-Identity-5GC & 12 & 00:15 & 1.6-4.8 & 0-10 \\ \hline

		Extend options in & EstablishmentCause: add new-code1, new-code2 & 35 & 00:17 & 0.4-3.1 & 0-10 \\ \hline

		Reduce options in & EstablishmentCause: remove mo-Signalling, mo-Data, delayTolerantAccess-v1020, mo-VoiceCall-v1280, spare1 & 12 & 00:15 & 1.3-3.1 & 0-10 \\ \hline

		Scramble options in & RRCConnectionRequest-r8 from "ue-Identity, establishmentCause, spare" to "spare, ue-Identity, establishmentCause" & 12 & 00:15 & 1.4-18.0 & 0-10 \\ \hline

		Change size constraints of & \makecell[cl]{randomValue from 40 bits to 80 bits; \\ m-TMSI from 32 bits to 72 bits} & 45 & 00:17 & 0.9-3.8 & 0-10 \\ \hline

		Combination of other strategies & \makecell[cl]{
		- Change data type: randomValue - BIT STRING to INTEGER \\
		- Change size: m-TMSI - 32 bits to 24 bits \\ 
		- Scramble in: RRCConnectionRequest-r8 \\ 
		- Reduced options: remove mo-Signalling, mo-Data, \\ \hspace{0.1cm} delayTolerantAccess-v1020, mo-VoiceCall-v1280, spare1 \\ 
		- Extend options in RRCConnectionRequest-r8: new-field of \\ \hspace{0.1cm} EstablishmentCause type
		}
		& 21 & 00:15 & 0.9-4.9 & 0-10 \\ \hline

		Perturbation\textsuperscript{a} & Send other UL-RRC messages in place of RRCConnectionRequest & 28 & 00:16 & 0.6-6.1 & 0-10 \\ \hline
		
		\multicolumn{6}{|l|}{\makecell[cl]{
		\textsuperscript{b}Uses non-mutated ASN.1 schema while generating other RRC messages and generates messages as identified in Table \ref{tab:rrc_messages_in_3gpp}}} \\ \hline
	\end{tabularx}
\end{table*}

\begin{table*}[ht!]
	\scriptsize
	\centering
	\caption{Results from generation-based fuzzing -- RRCConnectionSetup}
	\label{tab:results_generation_strategies_rrc_conn_set}
	\renewcommand{\tabularxcolumn}[1]{m{#1}}
	\newcolumntype{A}{>{\raggedright\hsize=1.1\hsize\linewidth=\hsize}X}
	\newcolumntype{B}{>{\raggedright\hsize=3.25\hsize\linewidth=\hsize}X}
	\newcolumntype{C}{>{\centering\arraybackslash\hsize=0.45\hsize\linewidth=\hsize}X}
	\newcolumntype{D}{>{\centering\arraybackslash\hsize=0.45\hsize\linewidth=\hsize}X}
	\newcolumntype{E}{>{\centering\arraybackslash\hsize=0.5\hsize\linewidth=\hsize}X}
	\newcolumntype{F}{>{\centering\arraybackslash\hsize=0.25\hsize\linewidth=\hsize}X}

	\begin{tabularx}{\textwidth}{|A|B|C|D|E|F|}
		\hline\textbf{ASN.1 specification mutation} & \textbf{Mutation selected} & \textbf{Number of RRC PDUs replaced} & \textbf{Time taken to test (hours, minutes)} & \textbf{Added time for fuzzing (milliseconds)} &\textbf{Seeds used} \\ \hline

		Change primitive data type of & RRC-TransactionIdentifier from INTEGER(0..3) to BIT STRING (SIZE (2)) & 11 & 00:16 & 1.55-3.11 & 0-10 \\ \hline

		Change structured types & RRCConnectionSetup-r8-IEs to RRCConnectionSetup-v8a0-IEs & 21 & 00:19 & 0.86-5.99 & 0-10 \\ \hline

		Extend options in & RRCConnectionSetup-r8-IEs: add new-field of type RRCConnectionSetup-v8a0-IEs		& 11 & 00:16 & 1.13-9.33 & 0-10 \\ \hline

		\multirow{2}{*}{Reduce options in} & RRCConnectionSetup-r8-IEs: remove RRCConnectionSetup-v8a0-IEs		& 19 & 00:19 & 0.80-5.56 & 0-10 \\ \cline{2-6}
		
		& criticialExtensions: remove criticalExtensionsFuture		& 11 & 00:16 & 1.59-3.42 & 0-10 \\ \hline

		Scramble options in & RRCConnectionSetup from "from rrc-TransactionIdentifier, criticalExtensions" to "criticalExtensions, rrc-TransactionIdentifier" & 11 & 00:16 & 0.78-4.94 & 0-10 \\ \hline
		
		Change size constraints of & RRC-TransactionIdentifier to INTEGER (0..31) & 15 & 00:19 & 1.16-5.07 & 0-10 \\ \hline

		Combination of other strategies & \makecell[cl]{
		- Change data type: RRC-TransactionIdentifier to BIT STRING \\\hspace{0.1cm}(SIZE (2))\\
		- Extend options: new-field of RRCConnectionSetup-v8a0-IEs\\
		- Scramble in: RRCConnectionSetup
		}
		& 11 & 00:16 & 0.83-2.97 & 0-10 \\ \hline

		Perturbation\textsuperscript{a} & Send other UL-RRC messages in place of RRCConnectionSetup & 10 & 00:16 & 1.75-9.87 & 0-10 \\ \hline
		
		\multicolumn{6}{|l|}{\makecell[cl]{\textsuperscript{a}Uses non-mutated ASN.1 schema while generating other RRC messages and generates messages as identified in Table \ref{tab:rrc_messages_in_3gpp}}} \\ \hline
	\end{tabularx}
\end{table*}

\begin{table*}[t!]
	\scriptsize
	\centering
	\caption{Results from generation-based fuzzing -- DLInformationTransfer}
	\label{tab:results_generation_strategies_dl_inf_trf}
	\renewcommand{\tabularxcolumn}[1]{m{#1}}
	\newcolumntype{A}{>{\raggedright\hsize=1.1\hsize\linewidth=\hsize}X}
	\newcolumntype{B}{>{\raggedright\hsize=3.25\hsize\linewidth=\hsize}X}
	\newcolumntype{C}{>{\centering\arraybackslash\hsize=0.45\hsize\linewidth=\hsize}X}
	\newcolumntype{D}{>{\centering\arraybackslash\hsize=0.45\hsize\linewidth=\hsize}X}
	\newcolumntype{E}{>{\centering\arraybackslash\hsize=0.5\hsize\linewidth=\hsize}X}
	\newcolumntype{F}{>{\centering\arraybackslash\hsize=0.25\hsize\linewidth=\hsize}X}

	\begin{tabularx}{\textwidth}{|A|B|C|D|E|F|}
		\hline\textbf{ASN.1 specification mutation} & \textbf{Mutation selected} & \textbf{Number of RRC PDUs replaced} & \textbf{Time taken to test (hours, minutes)} & \textbf{Added time for fuzzing (milliseconds)} &\textbf{Seeds used} \\ \hline

		\multirow{3}{*}{\makecell[cl]{Change primitive data \\type of}} & RRC-TransactionIdentifier from INTEGER(0..3) to BIT STRING (SIZE (2)) & 31 & 00:35 & 0.65-6.23 & 0-10 \\ \cline{2-6}

		& DedicatedInfoNAS from OCTET STRING to BIT STRING & 30 & 00:36 & 0.70-3.12 & 0-10 \\ \cline{2-6}

		& DedicatedInfoNAS OCTET STRING to INTEGER & 29 & 00:35 & 0.70-8.51 & 0-10 \\ \hline

		\multirow{2}{*}{\makecell[cl]{Change structured \\types}} & DLInformationTransfer-r8-IEs to DLInformationTransfer-v8a0-IEs & 30 & 00:36 & 0.70-4.96 & 0-10 \\ \cline{2-6}

		& DLInformationTransfer-r8-IEs to DLInformationTransfer-15-IEs & 30 & 00:33 & 0.69-3.91 & 0-10 \\ \hline

		Extend options in & DLInformationTransfer-r8-IEs: add new-field of type DLInformationTransfer-v8a0-IEs		& 30 & 00:34 & 0.63-3.40 & 0-10 \\ \hline

		Reduce options in & DLInformationTransfer-r8-IEs: remove DLInformationTransfer-v8a0-IEs		& 33 & 00:37 & 0.70-4.51 & 0-10 \\ \hline

		Scramble options in & DLInformationTransfer from "from rrc-TransactionIdentifier, criticalExtensions" to "criticalExtensions, rrc-TransactionIdentifier" & 30 & 00:36 & 0.76-3.42 & 0-10 \\ \hline
		
		\multirow{2}{*}{\makecell[cl]{Change size \\constraints of}} & dedicatedInfoNAS to OCTET STRING (SIZE 20) & 33 & 00:37 & 0.78-4.16 & 0-10 \\ \cline{2-6}

		& RRC-TransactionIdentifier to INTEGER (0..31) & 30 & 00:35 & 0.70-3.20 & 0-10 \\ \hline

		Combination of other strategies & \makecell[cl]{
		- Change data type: DedicatedInfoNAS to BIT STRING\\
		- Extend options: new-field of DLInformationTransfer-v8a0-IEs\\
		- Change size: RRC-TransactionIdentifier to INTEGER (0..31)\\
		- Scramble in: DLInformationTransfer
		}
		& 30 & 00:35 & 0.79-5.24 & 0-10 \\ \hline

		Perturbation\textsuperscript{a} & Send other UL-RRC messages in place of DLInformationTransfer & 26 & 00:32 & 0.66-5.26 & 0-10 \\ \hline
		
		\multicolumn{6}{|l|}{\makecell[cl]{\textsuperscript{a}Uses non-mutated ASN.1 schema while generating other RRC messages and generates messages as identified in Table \ref{tab:rrc_messages_in_3gpp}}} \\ \hline
	\end{tabularx}
\end{table*}

\begin{table*}[ht!]
	\scriptsize
	\centering
	\caption{Results from generation-based fuzzing -- RRCConnectionSetupComplete}
	\label{tab:results_generation_strategies_con_set_comp}
	\renewcommand{\tabularxcolumn}[1]{m{#1}}
	\newcolumntype{A}{>{\raggedright\hsize=1.1\hsize\linewidth=\hsize}X}
	\newcolumntype{B}{>{\raggedright\hsize=3.25\hsize\linewidth=\hsize}X}
	\newcolumntype{C}{>{\centering\arraybackslash\hsize=0.45\hsize\linewidth=\hsize}X}
	\newcolumntype{D}{>{\centering\arraybackslash\hsize=0.45\hsize\linewidth=\hsize}X}
	\newcolumntype{E}{>{\centering\arraybackslash\hsize=0.5\hsize\linewidth=\hsize}X}
	\newcolumntype{F}{>{\centering\arraybackslash\hsize=0.25\hsize\linewidth=\hsize}X}

	\begin{tabularx}{\textwidth}{|A|B|C|D|E|F|}
		\hline\textbf{ASN.1 specification mutation} & \textbf{Mutation selected} & \textbf{Number of RRC PDUs replaced} & \textbf{Time taken to test (hours, minutes)} & \textbf{Added time for fuzzing (milliseconds)} &\textbf{Seeds used} \\ \hline

		\multirow{6}{*}{\makecell[cl]{Change primitive data \\type of}} & RRC-TransactionIdentifier from INTEGER(0..3) to BIT STRING (SIZE (2)) & 12 & 00:45 & 0.61-3.01 & 0-10 \\ \cline{2-6}

		& selectedPLMN-Identity from INTEGER (1..maxPLMN-r11) to BIT STRING (SIZE (2)) & 11 & 00:37 & 0.94-91.68 & 0-10 \\ \cline{2-6}

		& DedicatedInfoNAS from OCTET STRING to BIT STRING & 17 & 00:35 & 0.64-12.23 & 0-10 \\ \cline{2-6}

		& DedicatedInfoNAS OCTET STRING to INTEGER & 10 & 00:16 & 1.20-2.93 & 0-10 \\ \hline

		\multirow{3}{*}{\makecell[cl]{Change structured \\types}} & RRCConnectionSetupComplete-r8-IEs to RegisteredMME & 11 & 00:17 & 0.87-3.08 & 0-10 \\ \cline{2-6}

		& RRCConnectionSetupComplete-r8-IEs to RRCConnectionSetupComplete-v8a0-IEs & 17 & 00:36 & 0.68-5.58 & 0-10 \\ \hline

		Extend options in & RRCConnectionSetupComplete-r8-IEs: add new-field of type RegisteredMME & 10 & 00:16 & 0.73-2.83 & 0-10 \\ \hline

		Reduce options in & RRCConnectionSetupComplete-r8-IEs: remove dedicatedInfoNAS & 18 & 00:37 & 0.62-11.35 & 0-10 \\ \hline

		Scramble options in & RRCConnectionSetupComplete-r8-IEs from "selectedPLMN-Identity, registeredMME, dedicatedInfoNAS, nonCriticalExtension" to "nonCriticalExtension, registeredMME, selectedPLMN-Identity, dedicatedInfoNAS" & 11 & 00:16 & 0.59-14.10 & 0-10 \\ \hline
		
		\multirow{3}{*}{\makecell[cl]{Change size \\constraints of}} & dedicatedInfoNAS to OCTET STRING (SIZE 20) & 17 & 00:36 & 0.58-3.41 & 0-10 \\ \cline{2-6}

		& selectedPLMN-Identity to INTEGER (1..31) & 15 & 00:38 & 0.63-11.88 & 0-10 \\ \cline{2-6}

		& RRC-TransactionIdentifier to INTEGER (0..1) & 19 & 00:33 & 0.65-3.19 & 0-10 \\ \hline

		Combination of other strategies & \makecell[cl]{
		- Change data type: DedicatedInfoNAS to BIT STRING\\
		- Extend options: new-field of RegisteredMME\\
		- Change size: selectedPLMN-Identity to INTEGER (1..31)\\
		- Scramble in: RRCConnectionSetupComplete-r8
		}
		& 19 & 00:37 & 0.68-9.82 & 0-10 \\ \hline

		Perturbation\textsuperscript{a} & Send other UL-RRC messages in place of RRCConnectionSetupComplete & 10 & 00:16 & 0.78-2.83 & 0-10 \\ \hline
		
		\multicolumn{6}{|l|}{\makecell[cl]{\textsuperscript{a}Uses non-mutated ASN.1 schema while generating other RRC messages and generates messages as identified in Table \ref{tab:rrc_messages_in_3gpp}}} \\ \hline
	\end{tabularx}
\end{table*}

\begin{table*}[ht!]
	\scriptsize
	\centering
	\caption{Results from generation-based fuzzing -- UECapabilityInformation}
	\label{tab:results_generation_strategies_ue_cap_inf}
	\renewcommand{\tabularxcolumn}[1]{m{#1}}
	\newcolumntype{A}{>{\raggedright\hsize=1.1\hsize\linewidth=\hsize}X}
	\newcolumntype{B}{>{\raggedright\hsize=3.25\hsize\linewidth=\hsize}X}
	\newcolumntype{C}{>{\centering\arraybackslash\hsize=0.45\hsize\linewidth=\hsize}X}
	\newcolumntype{D}{>{\centering\arraybackslash\hsize=0.45\hsize\linewidth=\hsize}X}
	\newcolumntype{E}{>{\centering\arraybackslash\hsize=0.5\hsize\linewidth=\hsize}X}
	\newcolumntype{F}{>{\centering\arraybackslash\hsize=0.25\hsize\linewidth=\hsize}X}

	\begin{tabularx}{\textwidth}{|A|B|C|D|E|F|}
		\hline\textbf{ASN.1 specification mutation} & \textbf{Mutation selected} & \textbf{Number of RRC PDUs replaced} & \textbf{Time taken to test (hours, minutes)} & \textbf{Added time for fuzzing (milliseconds)} &\textbf{Seeds used} \\ \hline

		\multirow{3}{*}{\makecell[cl]{Change primitive data \\type of}} & RRC-TransactionIdentifier to BIT STRING (SIZE (2)) & 44 & 00:29 & 0.66-30.18 & 0-10 \\ \cline{2-6}

		& ueCapabilityRAT-Container to BIT STRING & 45 & 00:17 & 0.48-3.40 & 0-10 \\ \cline{2-6}

		& ueCapabilityRAT-Container to INTEGER & 61 & 00:19 & 0.56-3.05 & 0-10 \\ \hline

		\multirow{2}{*}{\makecell[cl]{Change structured \\types}} & of UE-CapabilityRAT-ContainerList from \textit{"SEQUENCE (SIZE (0..maxRAT-Capabilities)) OF UE-CapabilityRAT-Container"} to \textit{"UE-CapabilityRAT-Container}" & 45 & 00:18 & 0.61-3.66 & 0-10 \\ \cline{2-6}

		& from UE-CapabilityRAT-ContainerList to UECapabilityInformation-v8a0-IEs & 60 & 00:17 & 0.64-3.38 & 0-10 \\ \hline

		\multirow{2}{*}{Extend options in} &  RAT-Type: add new-field1, new-field2, new-field3 & 49 & 00:17 & 0.49-3.79 & 0-10 \\ \cline{2-6}

		& UECapabilityInformation-r8-IEs: add new-field of type UE-CapabilityRAT-ContainerList & 54 & 00:21 & 0.55-8.74 & 0-10 \\ \hline

		\multirow{2}{*}{Reduce options in} & RAT-Type: remove geran-ps, cdma2000-1XRTT, nr, eutra-nr, spare1 & 49 & 00:17 & 0.60-4.13 & 0-10 \\ \cline{2-6}

		& UECapabilityInformation-r8-IEs: remove nonCriticalExtension & 53 & 00:21 & 0.54-3.22 & 0-10 \\ \hline

		\multirow{2}{*}{Scramble options in} & UECapabilityInformation-r8-IEs from "ue-CapabilityRAT-ContainerList, nonCriticalExtension" to "nonCriticalExtension, ue-CapabilityRAT-ContainerList" & 49 & 00:29 & 0.49-16.32 & 0-10 \\ \cline{2-6}

		& UE-CapabilityRAT-Container from "rat-Type, ueCapabilityRAT-Container" to "ueCapabilityRAT-Container, rat-Type" & 49 & 00:17 & 0.54-3.64 & 0-10 \\ \hline

		\multirow{2}{*}{\makecell[cl]{Change size \\constraints of}} & RRC-TransactionIdentifier to INTEGER (0..31) & 46 & 00:23 & 0.63-6.13 & 0-10 \\ \cline{2-6}

		& ueCapabilityRAT-Container to OCTET STRING (SIZE(20)) & 46 & 00:17 & 0.59-3.29 & 0-10 \\ \hline

		Combination of other strategies & \makecell[cl]{
		- Change data type: RRC-TransactionIdentifier to BIT STRING \\\hspace{0.1cm}(SIZE (2))\\
		- Extend options: new-field of UE-CapabilityRAT-ContainerList\\
		- Reduce options: remove geran-ps, cdma2000-1XRTT, nr, eutra-nr, \\\hspace{0.1cm}spare1\\
		- Scramble in: UECapabilityInformation-r8-IEs \\ 
		- Change size: ueCapabilityRAT-Container to OCTET STRING \\\hspace{0.1cm}(SIZE(20))
		}
		& 54 & 00:21 & 0.61-5.21 & 0-10 \\ \hline

		Perturbation\textsuperscript{a} & Send other UL-RRC messages in place of UECapabilityInformation & 11 & 00:16 & 0.51-4.30 & 0-10 \\ \hline
		
		\multicolumn{6}{|l|}{\makecell[cl]{\textsuperscript{a}Uses non-mutated ASN.1 schema while generating other RRC messages and generates messages as identified in Table \ref{tab:rrc_messages_in_3gpp}}} \\ 
				
		\hline
	\end{tabularx}
\end{table*}

\begin{table*}[ht!]
	\scriptsize
	\centering
	\caption{Results from generation-based fuzzing -- ULInformationTransfer}
	\label{tab:results_generation_strategies_ul_inf_trf}
	\renewcommand{\tabularxcolumn}[1]{m{#1}}
	\newcolumntype{A}{>{\raggedright\hsize=1.1\hsize\linewidth=\hsize}X}
	\newcolumntype{B}{>{\raggedright\hsize=3.25\hsize\linewidth=\hsize}X}
	\newcolumntype{C}{>{\centering\arraybackslash\hsize=0.45\hsize\linewidth=\hsize}X}
	\newcolumntype{D}{>{\centering\arraybackslash\hsize=0.45\hsize\linewidth=\hsize}X}
	\newcolumntype{E}{>{\centering\arraybackslash\hsize=0.5\hsize\linewidth=\hsize}X}
	\newcolumntype{F}{>{\centering\arraybackslash\hsize=0.25\hsize\linewidth=\hsize}X}

	\begin{tabularx}{\textwidth}{|A|B|C|D|E|F|}
		\hline\textbf{ASN.1 specification mutation} & \textbf{Mutation selected} & \textbf{Number of RRC PDUs replaced} & \textbf{Time taken to test (hours, minutes)} & \textbf{Added time for fuzzing (milliseconds)} &\textbf{Seeds used} \\ \hline

		\multirow{3}{*}{\makecell[cl]{Change primitive \\data type of \\DedicatedInfoNAS}} & from OCTET STRING to BIT STRING & 51 & 00:35 & 0.36-4.5 & 0-10 \\ \cline{2-6}

		& from OCTET STRING to INTEGER & 56 & 00:35 & 0.54-9.76 & 0-10 \\ \cline{2-6}

		& from OCTET STRING to C-RNTI (BIT STRING (SIZE (16))) & 55 & 00:38 & 0.60-7.42 & 0-10 \\ \hline

		Change structured types & from  ULInformationTransfer-r8-IEs to ULInformationTransfer-v8a0-IEs & 56 & 00:33 & 0.66-6.44 & 0-10 \\ \hline

		Extend options in & ULInformationTransfer-r8-IEs: add new-field of ULInformationTransfer-v8a0-IEs & 56 & 00:35 & 0.65-7.55 & 0-10 \\ \hline

		Reduce options in & ULInformationTransfer-r8-IEs: remove dedicatedInfoNAS & 55 & 00:38 & 0.65-6.96 & 0-10 \\ \hline

		Scramble options in & ULInformationTransfer-r8-IEs from "dedicatedInfoType, nonCriticalExtension" to "nonCriticalExtension, dedicatedInfoType" & 56 & 00:33 & 0.66-11.74 & 0-10 \\ \hline

		Change size constraints of & DedicatedInfoNAS to OCTET STRING (SIZE (20)) & 54 & 00:37 & 0.62-6.00 & 0-10 \\ \hline

		Combination of other strategies & \makecell[cl]{
		- Change data type: DedicatedInfoNAS to INTEGER\\
		- Extend options : new-field of ULInformationTransfer-v8a0-IEs\\
		- Scramble in: ULInformationTransfer-r8-IEs
		}
		& 55 & 00:38 & 0.72-8.32 & 0-10 \\ \hline

		Perturbation\textsuperscript{a} & Send other UL-RRC messages in place of ULInformationTransfer & 12 & 00:15 & 0.62-8.66 & 0-10 \\ \hline
		
		\multicolumn{6}{|l|}{\makecell[cl]{\textsuperscript{a}Uses non-mutated ASN.1 schema while generating other RRC messages and generates messages as identified in Table \ref{tab:rrc_messages_in_3gpp}}} \\ \hline
	\end{tabularx}
\end{table*}

\begin{table*}[ht!]
	\scriptsize
	\centering
	\caption{Results from generation-based fuzzing -- SecurityModeComplete}
	\label{tab:results_generation_strategies_sec_mode_comp}
	\renewcommand{\tabularxcolumn}[1]{m{#1}}
	\newcolumntype{A}{>{\raggedright\hsize=1.1\hsize\linewidth=\hsize}X}
	\newcolumntype{B}{>{\raggedright\hsize=3.25\hsize\linewidth=\hsize}X}
	\newcolumntype{C}{>{\centering\arraybackslash\hsize=0.45\hsize\linewidth=\hsize}X}
	\newcolumntype{D}{>{\centering\arraybackslash\hsize=0.45\hsize\linewidth=\hsize}X}
	\newcolumntype{E}{>{\centering\arraybackslash\hsize=0.5\hsize\linewidth=\hsize}X}
	\newcolumntype{F}{>{\centering\arraybackslash\hsize=0.25\hsize\linewidth=\hsize}X}

	\begin{tabularx}{\textwidth}{|A|B|C|D|E|F|}
		\hline\textbf{ASN.1 specification mutation} & \textbf{Mutation selected} & \textbf{Number of RRC PDUs replaced} & \textbf{Time taken to test (hours, minutes)} & \textbf{Added time for fuzzing (milliseconds)} &\textbf{Seeds used} \\ \hline

		Change primitive data type of & RRC-TransactionIdentifier from INTEGER(0..3) to BIT STRING (SIZE (2)) & 66 & 00:17 & 0.69-3.97 & 0-10 \\ \hline

		Change structured types & from SecurityModeComplete-r8-IEs to SecurityModeComplete-v8a0-IEs & 68 & 00:14 & 0.56-4.41 & 0-10 \\ \hline

		Extend options in & SecurityModeComplete-r8-IEs: add new-field of type SecurityModeComplete-v8a0-IEs & 66 & 00:17 & 0.58-3.77 & 0-10 \\ \hline

		\multirow{2}{*}{Reduce options in} & SecurityModeComplete-r8-IEs: remove SecurityModeComplete-v8a0-IEs & 34 & 00:28 & 0.39-4.59 & 0-10 \\ \cline{2-6}

		& criticialExtensions: remove criticalExtensionsFuture & 33 & 00:38 & 0.71-4.68 & 0-10 \\ \hline

		Scramble options in & SecurityModeComplete from "rrc-TransactionIdentifier, criticalExtensions" to "criticalExtensions, rrc-TransactionIdentifier" & 45 & 00:16 & 0.59-4.41 & 0-10 \\ \hline

		Change size constraints of & RRC-TransactionIdentifier to INTEGER (0..31) & 59 & 00:21 & 0.65-4.04 & 0-10 \\ \hline

		Combination of other strategies & \makecell[cl]{
		- Change data type: RRC-TransactionIdentifier to BIT STRING \\\hspace{0.1cm}(SIZE (2))\\
		- Extend options: new-field of SecurityModeComplete-v8a0-IEs\\
		- Scramble in: SecurityModeComplete
		}
		& 62 & 00:17 & 0.37-5.03 & 0-10 \\ \hline

		Perturbation\textsuperscript{a} & Send other UL-RRC messages in place of SecurityModeComplete & 30 & 00:36 & 0.47-3.03 & 0-10 \\ \hline
		
		\multicolumn{6}{|l|}{\makecell[cl]{\textsuperscript{a}Uses non-mutated ASN.1 schema while generating other RRC messages and generates messages as identified in Table \ref{tab:rrc_messages_in_3gpp}}} \\ \hline
	\end{tabularx}
\end{table*}

\begin{table*}[t!]
	\scriptsize
	\centering
	\caption{Results from generation-based fuzzing -- UECapabilityEnquiry}
	\label{tab:results_generation_strategies_ue_cap_enq}
	\renewcommand{\tabularxcolumn}[1]{m{#1}}
	\newcolumntype{A}{>{\raggedright\hsize=1.1\hsize\linewidth=\hsize}X}
	\newcolumntype{B}{>{\raggedright\hsize=3.25\hsize\linewidth=\hsize}X}
	\newcolumntype{C}{>{\centering\arraybackslash\hsize=0.45\hsize\linewidth=\hsize}X}
	\newcolumntype{D}{>{\centering\arraybackslash\hsize=0.45\hsize\linewidth=\hsize}X}
	\newcolumntype{E}{>{\centering\arraybackslash\hsize=0.5\hsize\linewidth=\hsize}X}
	\newcolumntype{F}{>{\centering\arraybackslash\hsize=0.25\hsize\linewidth=\hsize}X}

	\begin{tabularx}{\textwidth}{|A|B|C|D|E|F|}
		\hline\textbf{ASN.1 specification mutation} & \textbf{Mutation selected} & \textbf{Number of RRC PDUs replaced} & \textbf{Time taken to test (hours, minutes)} & \textbf{Added time for fuzzing (milliseconds)} &\textbf{Seeds used} \\ \hline

		\multirow{2}{*}{\makecell[cl]{Change primitive data \\type of}} & RRC-TransactionIdentifier from INTEGER(0..3) to BIT STRING (SIZE (2)) & 61 & 00:19 & 0.60-3.19 & 0-10 \\ \cline{2-6}

		& RAT-Type from ENUMERATED to OCTET STRING (SIZE (5))		& 29 & 00:18 & 0.91-2.49 & 0-10 \\ \hline

		Change structured types & UECapabilityEnquiry-r8-IEs to UECapabilityEnquiry-v8a0-IEs & 40 & 00:25 & 0.66-3.60 & 0-10 \\ \hline

		\multirow{2}{*}{Extend options in} & UECapabilityEnquiry-r8-IEs: add new-field of type UECapabilityEnquiry-v8a0-IEs		& 45 & 00:15 & 0.56-4.00 & 0-10 \\ \cline{2-6}
		
		& RAT-Type: add new-field1, new-field2		& 50 & 00:23 & 0.65-4.57 & 0-10 \\ \hline

		\multirow{2}{*}{Reduce options in} & UECapabilityEnquiry-r8-IEs: remove UECapabilityEnquiry-v8a0-IEs		& 49 & 00:23 & 0.57-3.63 & 0-10 \\ \cline{2-6}
		
		& RAT-Type: remove cdma2000-1XRTT, nr, eutra-nr, spare1		& 42 & 00:17 & 0.59-4.12 & 0-10 \\ \hline

		Scramble options in & UECapabilityEnquiry from "from rrc-TransactionIdentifier, criticalExtensions" to "criticalExtensions, rrc-TransactionIdentifier" & 42 & 00:22 & 0.63-4.26 & 0-10 \\ \hline
		
		Change size constraints of & RRC-TransactionIdentifier to INTEGER (0..31) & 42 & 00:20 & 0.60-3.11 & 0-10 \\ \hline

		Combination of other strategies & \makecell[cl]{
		- Change data type: RAT-Type to OCTET STRING (SIZE (5))\\
		- Extend options: new-field of UECapabilityEnquiry-v8a0-IEs\\
		- Change size: RRC-TransactionIdentifier to INTEGER (0..31)\\
		- Scramble in: UECapabilityEnquiry
		}
		& 40 & 00:18 & 0.55-2.33 & 0-10 \\ \hline

		Perturbation\textsuperscript{a} & Send other UL-RRC messages in place of UECapabilityEnquiry & 27 & 00:32 & 0.74-5.94 & 0-10 \\ \hline
		
		\multicolumn{6}{|l|}{\makecell[cl]{\textsuperscript{a}Uses non-mutated ASN.1 schema while generating other RRC messages and generates messages as identified in Table \ref{tab:rrc_messages_in_3gpp}}} \\ \hline
	\end{tabularx}
\end{table*}

\begin{table*}[t!]
	\scriptsize
	\centering
	\caption{Results from generation-based fuzzing -- RRCConnectionReconfiguration}
	\label{tab:results_generation_strategies_rrc_conn_reconf}
	\renewcommand{\tabularxcolumn}[1]{m{#1}}
	\newcolumntype{A}{>{\raggedright\hsize=1.1\hsize\linewidth=\hsize}X}
	\newcolumntype{B}{>{\raggedright\hsize=3.25\hsize\linewidth=\hsize}X}
	\newcolumntype{C}{>{\centering\arraybackslash\hsize=0.45\hsize\linewidth=\hsize}X}
	\newcolumntype{D}{>{\centering\arraybackslash\hsize=0.45\hsize\linewidth=\hsize}X}
	\newcolumntype{E}{>{\centering\arraybackslash\hsize=0.5\hsize\linewidth=\hsize}X}
	\newcolumntype{F}{>{\centering\arraybackslash\hsize=0.25\hsize\linewidth=\hsize}X}

	\begin{tabularx}{\textwidth}{|A|B|C|D|E|F|}
		\hline\textbf{ASN.1 specification mutation} & \textbf{Mutation selected} & \textbf{\# of PDUs replaced} & \textbf{Time to test (hr, min)} & \textbf{Added time (ms)} &\textbf{Seeds used} \\ \hline

		\multirow{3}{*}{\makecell[cl]{Change primitive data \\type of}} & RRC-TransactionIdentifier from INTEGER(0..3) to BIT STRING (SIZE (2)) & 35 & 00:37 & 0.73-4.20 & 0-10 \\ \cline{2-6}

		& DedicatedInfoNAS from OCTET STRING to BIT STRING & 50 & 00:27 & 0.64-4.22 & 0-10 \\ \cline{2-6}

		& DedicatedInfoNAS OCTET STRING to INTEGER & 33 & 00:38 & 0.73-17.13 & 0-10 \\ \hline

		Change structured types & RRCConnectionReconfiguration-r8-IEs to RRCConnectionReconfiguration-v890-IEs & 35 & 00:37 & 0.69-2.94 & 0-10 \\ \hline

		Extend options in & RRCConnectionReconfiguration-r8-IEs: add new-field of type RRCConnectionReconfiguration-v890-IEs		& 35 & 00:37 & 0.68-3.96 & 0-10 \\ \hline

		Reduce options in & RRCConnectionReconfiguration-r8-IEs: remove RRCConnectionReconfiguration-v890-IEs		& 34 & 00:37 & 0.77-29.01 & 0-10 \\ \hline

		Scramble options in & RRCConnectionReconfiguration from "from rrc-TransactionIdentifier, criticalExtensions" to "criticalExtensions, rrc-TransactionIdentifier" & 55 & 00:27 & 0.61-2.45 & 0-10 \\ \hline
		
		Change size constraints of & RRC-TransactionIdentifier to INTEGER (0..31) & 35 & 00:37 & 0.72-5.11 & 0-10 \\ \hline

		Combination of other strategies & \makecell[cl]{
		- Change data type: DedicatedInfoNAS to BIT STRING\\
		- Extend options: new-field of \\\hspace{0.1cm}RRCConnectionReconfiguration-v890-IEs\\
		- Change size: RRC-TransactionIdentifier to INTEGER (0..31)\\
		- Scramble in: RRCConnectionReconfiguration
		}
		& 35 & 00:37 & 0.73-4.27 & 0-10 \\ \hline

		Perturbation\textsuperscript{a} & Send other UL-RRC messages in place of RRCConnectionReconfiguration & 44 & 00:25 & 0.56-8.73 & 0-10 \\ \hline
		
		\multicolumn{6}{|l|}{\makecell[cl]{\textsuperscript{a}Uses non-mutated ASN.1 schema while generating other RRC messages and generates messages as identified in Table \ref{tab:rrc_messages_in_3gpp}}} \\ \hline
	\end{tabularx}
\end{table*}

\begin{table*}[t!]
	\scriptsize
	\centering
	\caption{Results from generation-based fuzzing -- RRCConnectionReconfigurationComplete}
	\label{tab:results_generation_strategies_con_reconf_comp}
	\renewcommand{\tabularxcolumn}[1]{m{#1}}
	\newcolumntype{A}{>{\raggedright\hsize=1.1\hsize\linewidth=\hsize}X}
	\newcolumntype{B}{>{\raggedright\hsize=3.25\hsize\linewidth=\hsize}X}
	\newcolumntype{C}{>{\centering\arraybackslash\hsize=0.45\hsize\linewidth=\hsize}X}
	\newcolumntype{D}{>{\centering\arraybackslash\hsize=0.45\hsize\linewidth=\hsize}X}
	\newcolumntype{E}{>{\centering\arraybackslash\hsize=0.5\hsize\linewidth=\hsize}X}
	\newcolumntype{F}{>{\centering\arraybackslash\hsize=0.25\hsize\linewidth=\hsize}X}

	\begin{tabularx}{\textwidth}{|A|B|C|D|E|F|}
		\hline\textbf{ASN.1 specification mutation} & \textbf{Mutation selected} & \textbf{\# of PDUs replaced} & \textbf{Time to test (hr, min)} & \textbf{Added time (ms)} &\textbf{Seeds used} \\ \hline

		Change primitive data type of & RRC-TransactionIdentifier from INTEGER(0..3) to BIT STRING (SIZE (2)) & 54 & 00:27 & 0.52-4.90 & 0-10 \\ \hline

		Change structured types & from  RRCConnectionReconfigurationComplete-r8-IEs to RRCConnectionReconfigurationComplete-v8a0-IEs & 61 & 00:25 & 0.55-5.17 & 0-10 \\ \hline

		Extend options in & RRCConnectionReconfigurationComplete-r8-IEs: add new-field of type RRCConnectionReconfigurationComplete-v8a0-IEs & 55 & 00:23 & 0.67-4.18 & 0-10 \\ \hline

		\multirow{2}{*}{Reduce options in} & RRCConnectionReconfigurationComplete-r8-IEs: remove RRCConnectionReconfigurationComplete-v8a0-IEs & 63 & 00:29 & 0.60-3.82 & 0-10 \\ \cline{2-6}

		& criticialExtensions: remove criticalExtensionsFuture & 50 & 00:26 & 0.50-4.68 & 0-10 \\ \hline

		Scramble options in & RRCConnectionReconfigurationComplete from "rrc-TransactionIdentifier, criticalExtensions" to "criticalExtensions, rrc-TransactionIdentifier" & 65 & 00:21 & 0.65-5.61 & 0-10 \\ \hline

		Change size constraints of & RRC-TransactionIdentifier from INTEGER (0..3) to INTEGER (0..31) & 50 & 00:23 & 0.55-3.70 & 0-10 \\ \hline

		Combination of other strategies & \makecell[cl]{
		- Change data type: RRC-TransactionIdentifier to BIT STRING \\\hspace{0.1cm}(SIZE (2))\\
		- Extend options: new-field of \\\hspace{0.1cm}RRCConnectionReconfigurationComplete-v8a0-IEs\\
		- Scramble in: RRCConnectionReconfigurationComplete
		}
		& 65 & 00:21 & 0.65-4.77 & 0-10 \\ \hline

		Perturbation\textsuperscript{a} & Send other UL-RRC messages in place of RRCConnectionReconfigurationComplete & 52 & 00:27 & 0.48-20.10 & 0-10 \\ \hline
		
		\multicolumn{6}{|l|}{\makecell[cl]{\textsuperscript{a}Uses non-mutated ASN.1 schema while generating other RRC messages and generates messages as identified in Table \ref{tab:rrc_messages_in_3gpp}}} \\ \hline
	\end{tabularx}
\end{table*}

\begin{table*}[t!]
	\scriptsize
	\centering
	\caption{Results from generation-based fuzzing -- RRCConnectionRelease}
	\label{tab:results_generation_strategies_rrc_conn_rel}
	\renewcommand{\tabularxcolumn}[1]{m{#1}}
	\newcolumntype{A}{>{\raggedright\hsize=1.1\hsize\linewidth=\hsize}X}
	\newcolumntype{B}{>{\raggedright\hsize=3.25\hsize\linewidth=\hsize}X}
	\newcolumntype{C}{>{\centering\arraybackslash\hsize=0.45\hsize\linewidth=\hsize}X}
	\newcolumntype{D}{>{\centering\arraybackslash\hsize=0.45\hsize\linewidth=\hsize}X}
	\newcolumntype{E}{>{\centering\arraybackslash\hsize=0.5\hsize\linewidth=\hsize}X}
	\newcolumntype{F}{>{\centering\arraybackslash\hsize=0.25\hsize\linewidth=\hsize}X}

	\begin{tabularx}{\textwidth}{|A|B|C|D|E|F|}
		\hline\textbf{ASN.1 specification mutation} & \textbf{Mutation selected} & \textbf{\# of PDUs replaced} & \textbf{Time to test (hr, min)} & \textbf{Added time (ms)} &\textbf{Seeds used} \\ \hline

		\multirow{2}{*}{\makecell[cl]{Change primitive data \\type of}} & RRC-TransactionIdentifier from INTEGER(0..3) to BIT STRING (SIZE (2)) & 56 & 00:14 & 0.86-8.16 & 0-10 \\ \cline{2-6}

		& ReleaseCause from ENUMERATED to OCTET STRING (SIZE (5))		& 65 & 00:17 & 0.68-7.43 & 0-10 \\ \hline

		Change structured types & RRCConnectionRelease-r8-IEs to RRCConnectionRelease-v890-IEs & 51 & 00:14 & 0.68-4.57 & 0-10 \\ \hline

		\multirow{2}{*}{Extend options in} & RRCConnectionRelease-r8-IEs: add new-field of type RRCConnectionRelease-v890-IEs		& 56 & 00:20 & 0.66-4.32 & 0-10 \\ \cline{2-6}
		
		& ReleaseCause: add new-field1, new-field2		& 62 & 00:18 & 0.70-4.49 & 0-10 \\ \hline

		\multirow{2}{*}{Reduce options in} & RRCConnectionRelease-r8-IEs: remove RRCConnectionRelease-v890-IEs		& 57 & 00:19 & 0.60-4.52 & 0-10 \\ \cline{2-6}
		
		& ReleaseCause: remove cs-FallbackHighPriority-v1020, rrc-Suspend-v1320		& 56 & 00:14 & 0.72-7.83 & 0-10 \\ \hline

		Scramble options in & RRCConnectionRelease from "from rrc-TransactionIdentifier, criticalExtensions" to "criticalExtensions, rrc-TransactionIdentifier" & 42 & 00:21 & 0.56-7.95 & 0-10 \\ \hline
		
		Change size constraints of & RRC-TransactionIdentifier to INTEGER (0..31) & 37 & 00:19 & 0.66-5.20 & 0-10 \\ \hline

		Combination of other strategies & \makecell[cl]{
		- Change data type: ReleaseCause to OCTET STRING (SIZE (5))\\
		- Extend options: new-field of RRCConnectionRelease-v890-IEs\\
		- Change size: RRC-TransactionIdentifier to INTEGER (0..31)\\
		- Scramble in: RRCConnectionRelease
		}
		& 34 & 00:23 & 0.59-4.75 & 0-10 \\ \hline

		Perturbation\textsuperscript{a} & Send other UL-RRC messages in place of RRCConnectionRelease & 18 & 00:20 & 0.62-4.66 & 0-10 \\ \hline
		
		\multicolumn{6}{|l|}{\makecell[cl]{\textsuperscript{a}Uses non-mutated ASN.1 schema while generating other RRC messages and generates messages as identified in Table \ref{tab:rrc_messages_in_3gpp}}} \\ \hline
	\end{tabularx}
\end{table*}

\begin{table*}[t!]
	\scriptsize
	\centering
	\caption{Results from generation-based fuzzing -- SecurityModeCommand}
	\label{tab:results_generation_strategies_sec_mode_cmd}
	\renewcommand{\tabularxcolumn}[1]{m{#1}}
	\newcolumntype{A}{>{\raggedright\hsize=1.1\hsize\linewidth=\hsize}X}
	\newcolumntype{B}{>{\raggedright\hsize=3.25\hsize\linewidth=\hsize}X}
	\newcolumntype{C}{>{\centering\arraybackslash\hsize=0.45\hsize\linewidth=\hsize}X}
	\newcolumntype{D}{>{\centering\arraybackslash\hsize=0.45\hsize\linewidth=\hsize}X}
	\newcolumntype{E}{>{\centering\arraybackslash\hsize=0.5\hsize\linewidth=\hsize}X}
	\newcolumntype{F}{>{\centering\arraybackslash\hsize=0.25\hsize\linewidth=\hsize}X}

	\begin{tabularx}{\textwidth}{|A|B|C|D|E|F|}
		\hline\textbf{ASN.1 specification mutation} & \textbf{Mutation selected} & \textbf{\# of PDUs replaced} & \textbf{Time to test (hr, min)} & \textbf{Added time (ms)} &\textbf{Seeds used} \\ \hline

		\multirow{3}{*}{\makecell[cl]{Change primitive data \\type of}} & RRC-TransactionIdentifier from INTEGER(0..3) to BIT STRING (SIZE (2)) & 11 & 00:16 & 0.77-3.05 & 0-10 \\ \cline{2-6}

		& integrityProtAlgorithm from ENUMERATED to OCTET STRING (SIZE (5))		& 11 & 00:19 & 0.69-2.78 & 0-10 \\ \hline

		Change structured types & SecurityModeCommand-r8-IEs to SecurityModeCommand-v8a0-IEs & 11 & 00:16 & 0.77-1.39 & 0-10 \\ \hline

		\multirow{3}{*}{Extend options in} & SecurityModeCommand-r8-IEs: add new-field of type SecurityModeCommand-v8a0-IEs		& 11 & 00:19 & 0.54-3.17 & 0-10 \\ \cline{2-6}
		
		& integrityProtAlgorithm: add new-field1, new-field2		& 12 & 00:17 & 0.77-2.47 & 0-10 \\ \hline

		\multirow{4}{*}{Reduce options in} & SecurityModeCommand-r8-IEs: remove SecurityModeCommand-v8a0-IEs		& 12 & 00:19 & 0.68-3.92 & 0-10 \\ \cline{2-6}
		
		& CipheringAlgorithm-r12: remove eea3-v1130, spare4, spare3, spare2, spare1		& 12 & 00:20 & 0.70-2.78 & 0-10 \\ \hline

		Scramble options in & SecurityModeCommand from "from rrc-TransactionIdentifier, criticalExtensions" to "criticalExtensions, rrc-TransactionIdentifier" & 10 & 00:17 & 0.79-3.19 & 0-10 \\ \hline
		
		Change size constraints of & RRC-TransactionIdentifier to INTEGER (0..31) & 11 & 00:19 & 0.64-2.97 & 0-10 \\ \hline

		Combination of other strategies & \makecell[cl]{
		- Change data type: integrityProtAlgorithm to OCTET STRING \\\hspace{0.1cm}(SIZE (5))\\
		- Extend options: new-field of SecurityModeCommand-v8a0-IEs\\
		- Reduced options: remove eea3-v1130, spare4, spare3, spare2, \\\hspace{0.1cm}spare1\\
		- Change size: RRC-TransactionIdentifier to INTEGER (0..31)\\
		- Scramble in: SecurityModeCommand
		}
		& 11 & 00:16 & 0.72-3.40 & 0-10 \\ \hline

		Perturbation\textsuperscript{a} & Send other UL-RRC messages in place of SecurityModeCommand & 11 & 00:19 & 0.61-2.12 & 0-10 \\ \hline
		
		\multicolumn{6}{|l|}{\makecell[cl]{\textsuperscript{a}Uses non-mutated ASN.1 schema while generating other RRC messages and generates messages as identified in Table \ref{tab:rrc_messages_in_3gpp}}} \\ \hline
	\end{tabularx}
\end{table*}

\newcommand{\cmark}{\ding{52}}
\newcommand{\xmark}{\ding{56}}
\begin{table*}[ht!]
	\scriptsize
	\centering
	\caption{RRC messages from 3GPP TS 36.331 version 15.4.0 that are covered by Berserker (G=Generation-based, M=Mutation-based fuzzing; broadcast not included in downlink)}
	\label{tab:rrc_messages_in_3gpp}
	\setlength\tabcolsep{1.5pt}

	\begin{tabular}[]{l|lllllllllllllllllllllllllllllllllll|llllllllllllllllllllll}		
				
		\rotatebox{90}{\textbf{Uplink RRC messages}} & 
		\rotatebox{90}{RRCConnectionRequest} & 
		\rotatebox{90}{RRCConnectionSetupComplete} & 
		\rotatebox{90}{SecurityModeComplete} & 
		\rotatebox{90}{UECapabilityInformation} & 
		\rotatebox{90}{ULInformationTransfer} & 
		\rotatebox{90}{RRCConnectionReconfigurationComplete} & 
		\rotatebox{90}{CounterCheckResponse} & 
		\rotatebox{90}{CSFBParametersRequestCDMA2000} & 
		\rotatebox{90}{FailureInformation-r15} & 
		\rotatebox{90}{InDeviceCoexIndication-r11} & 
		\rotatebox{90}{InterFreqRSTDMeasurementIndication-r10} & 
		\rotatebox{90}{MBMSCountingResponse-r10} & 
		\rotatebox{90}{MBMSInterestIndication-r11} & 
		\rotatebox{90}{MeasReportAppLayer-r15} & 
		\rotatebox{90}{MeasurementReport} & 
		\rotatebox{90}{ProximityIndication-r9} & 
		\rotatebox{90}{RNReconfigurationComplete-r10} & 		
		\rotatebox{90}{RRCConnectionReestablishmentComplete} & 
		\rotatebox{90}{RRCConnectionReestablishmentRequest} & 	
		\rotatebox{90}{RRCConnectionResumeComplete-r13} & 
		\rotatebox{90}{RRCConnectionResumeRequest-r13} & 
		\rotatebox{90}{RRCEarlyDataRequest-r15} & 
		\rotatebox{90}{SCGFailureInformation-r12} & 
		\rotatebox{90}{SCGFailureInformationNR-r15} & 
		\rotatebox{90}{SecurityModeFailure} & 
		\rotatebox{90}{SidelinkUEInformation-r12} & 
		\rotatebox{90}{UEAssistanceInformation-r11} & 
		\rotatebox{90}{UEInformationResponse-r9} & 
		\rotatebox{90}{ULHandoverPreparationTransfer} & 
		\rotatebox{90}{ULInformationTransferMRDC-r15} & 
		\rotatebox{90}{WLANConnectionStatusReport-r13} &
		
		&&&

		
		\rotatebox{90}{\textbf{Downlink RRC messages}} & 
		\rotatebox{90}{RRCConnectionSetup} &
		\rotatebox{90}{SecurityModeCommand} &
		\rotatebox{90}{UECapabilityEnquiry} &
		\rotatebox{90}{RRCConnectionReconfiguration} &
		\rotatebox{90}{DLInformationTransfer} &
		\rotatebox{90}{RRCConnectionRelease} &
		\rotatebox{90}{CounterCheck} &
		\rotatebox{90}{CSFBParametersResponseCDMA2000} &
		\rotatebox{90}{HandoverFromEUTRAPreparationRequest} &
		\rotatebox{90}{LoggedMeasurementConfiguration-r10} &
		\rotatebox{90}{MobilityFromEUTRACommand} &
		\rotatebox{90}{RNReconfiguration-r10} &
		\rotatebox{90}{RRCConnectionReestablishment} &
		\rotatebox{90}{RRCConnectionReestablishmentReject} &
		\rotatebox{90}{RRCConnectionReject} &
		\rotatebox{90}{RRCConnectionResume-r13} &
		\rotatebox{90}{RRCEarlyDataComplete-r15} &
		\rotatebox{90}{UEInformationRequest-r9}
		\\ \hline

		\rotatebox{90}{\textbf{M}} &
		\cmark &
		\cmark &
		\cmark &
		\cmark &
		\cmark &
		\cmark &
		\xmark &
		\xmark &
		\xmark &
		\xmark &
		\xmark &
		\xmark &
		\xmark &
		\xmark &
		\xmark &
		\xmark &
		\xmark &
		\xmark &
		\xmark &
		\xmark &
		\xmark &
		\xmark &
		\xmark &
		\xmark &
		\xmark &
		\xmark &
		\xmark &
		\xmark &
		\xmark &
		\xmark &
		\xmark &

		&&&
		\rotatebox{90}{\textbf{M}} &		
		\cmark &
		\cmark &
		\cmark &
		\cmark &
		\cmark &
		\cmark &		
		\xmark &
		\xmark &
		\xmark &
		\xmark &
		\xmark &
		\xmark &
		\xmark &
		\xmark &
		\xmark &
		\xmark &
		\xmark &
		\xmark 	

		\\ \hline
		\rotatebox{90}{\textbf{G}} &
		\cmark &
		\cmark &
		\cmark &
		\cmark &
		\cmark &
		\cmark &
		\cmark &
		\cmark &
		\cmark &
		\cmark &
		\cmark &
		\cmark &
		\cmark &
		\cmark &
		\cmark &
		\cmark &
		\cmark &
		\cmark &
		\cmark &
		\cmark &
		\cmark &
		\cmark &
		\cmark &
		\cmark &
		\cmark &
		\cmark &
		\cmark &
		\cmark &
		\cmark &
		\cmark &
		\cmark &

		&&&
		\rotatebox{90}{\textbf{G}} &
		\cmark &
		\cmark &
		\cmark &
		\cmark &
		\cmark &
		\cmark &
		\cmark &
		\cmark &
		\cmark &
		\cmark &
		\cmark &
		\cmark &
		\cmark &
		\cmark &
		\cmark &
		\cmark &
		\cmark &
		\cmark
		\\ \hline
	\end{tabular}
\end{table*}

\begin{figure}[]
	\centering
	\includegraphics[width=10cm]{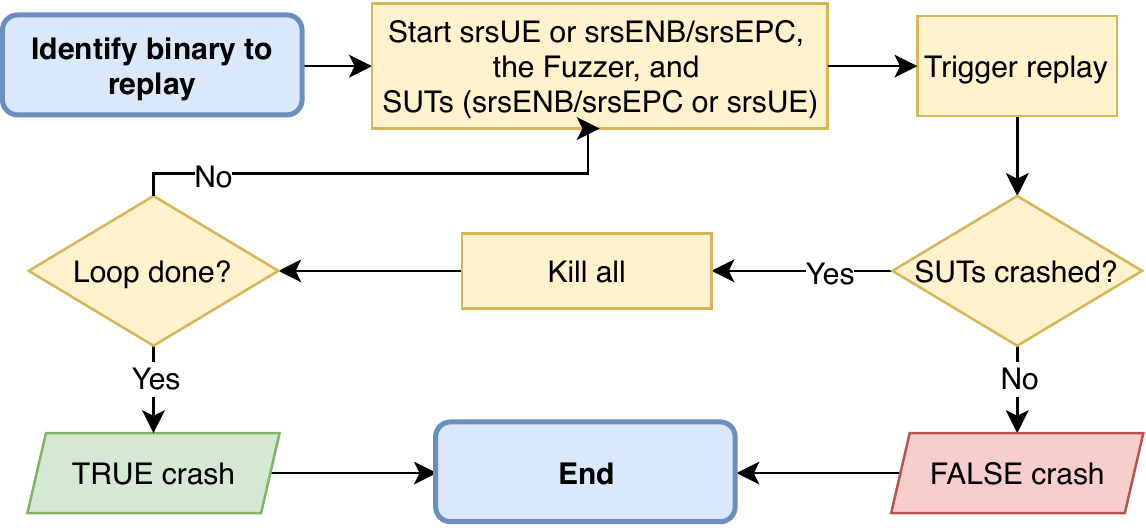}
	\caption {Flowchart for verifying a true crash.} 
	\label{fig:remove_false_positives} 
\end{figure}

\clearpage

\textbf{Filtering out false positives.} 
The Driver identifies unexpected behavior -- becoming unresponsive or crashing -- of the SUTs through their logs. The srsLTE application logs have a line "\textit{srsLTE crashed... backtrace saved in $'$./srsLTE.backtrace.crash$'$...}" when the application crashes. But all the unexpected behaviors are not necessarily caused by the Fuzzer; some may be caused by artifacts of the applications themselves. In order to filter out false positives, the Driver has an automated replaying mechanism to confirm the unexpected behavior. As shown in Figure \ref{fig:remove_false_positives}, the Driver replays all fuzzed messages (RRC PDU* binaries stored earlier) corresponding to the identified unexpected behaviors. An unexpected behavior is classified as true positive only if it is reproducible multiple times across multiple machines.
\section{Findings and Limitations} \label{findings_limitations}

\textbf{Two vulnerabilities in srsEPC.} 
Berserker found two previously unknown vulnerabilities both of which cause the srsEPC -- the CN part of srsLTE -- to crash. One of them was found through mutation-based uplink fuzzing and another through generation-based uplink fuzzing. Because of the sensitive nature of telecom networks, we do not disclose the exact byte sequence or message names in this paper. 

\begin{figure}[ht!]
	\centering
	\includegraphics[width=10cm]{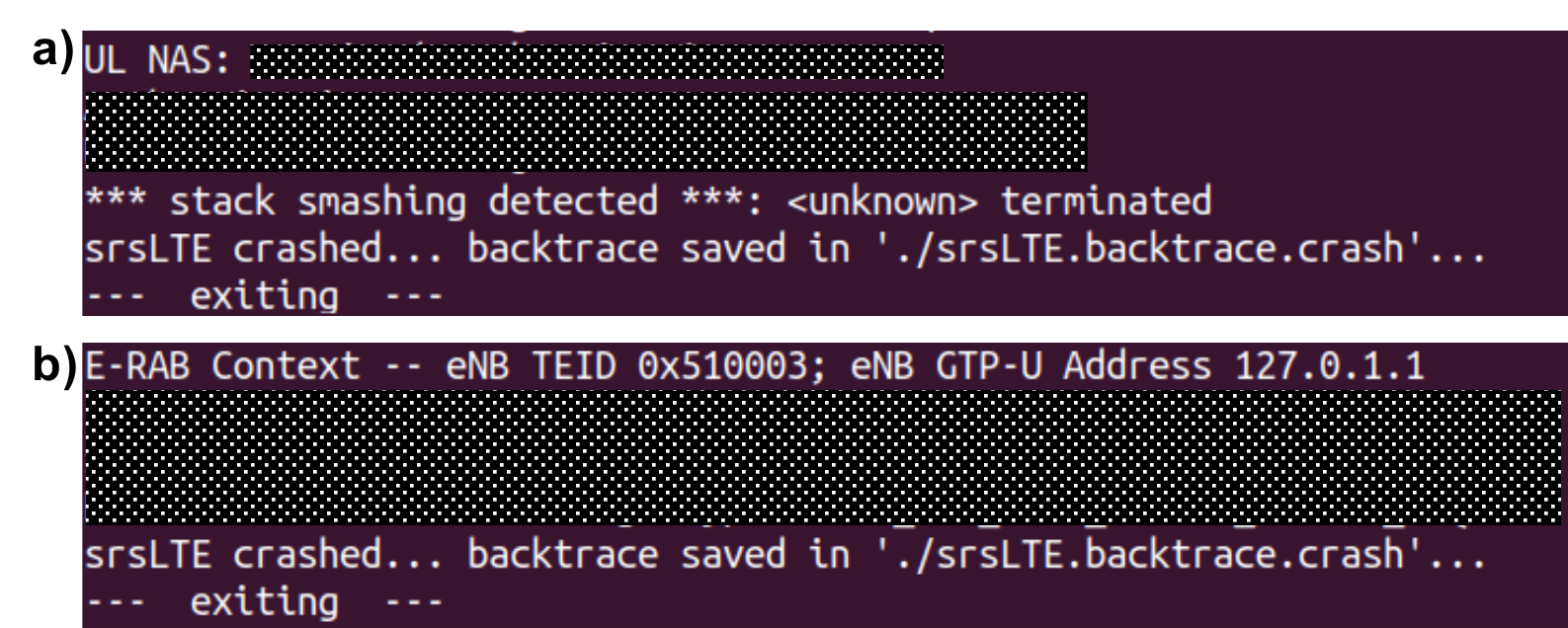}
	\caption{Crashes in srsEPC due to: a) improper parsing of NAS message; b) improper security processing of NAS message.}
	\label{fig:srs_epc_crash} 	
\end{figure}

The root cause of one vulnerability is buffer overflow because of improper parsing of a NAS message. It is triggered by a \textit{fuzzed} RRC message (RRC PDU* from \textbf{mutation}-based fuzzing) that tunneled a NAS message. The NAS message terminates at the srsEPC and causes the crash as shown in Figure \ref{fig:srs_epc_crash}.a. The \textit{stack smashing detected} error originates from safety mechanism against buffer overflow in C++ applications compiled using gcc \cite{gcc_smash}.

The other vulnerability's root cause is invalid memory address access because of improper security processing of a NAS message. It is triggered by a \textit{replaced} RRC message (RRC PDU* from \textbf{generation}-based fuzzing) that causes the srsEPC to crash as shown in Figure \ref{fig:srs_epc_crash}.b.

Because of the sensitive nature of telecom networks, we do not disclose the exact byte sequences or message names in this paper. But full details could be made available via Ericsson PSIRT.

\textbf{One vulnerability in openLTE.}
We investigated further and learned that srsLTE uses some lines of codes from openLTE \cite{openlte_2019} (another open-source LTE project) for parsing NAS messages. We fed the fuzzed RRC message, that caused buffer overflow in srsLTE, into openLTE ({\footnotesize \texttt{a5a66ed660e6094a9f50f7e00a0e9805dfbac724}} commit tag, 30th Jul 2017 – the latest commit at the time of writing) and discovered that the openLTE crashes too. This suggests that the buffer overflow vulnerability in srsLTE came from openLTE's code base.

\textbf{Effect in srsENB and srsUE.} In our experiments over the total period of $\sim$26 days, Berserker did not identify any fuzzed RRC message that crashes srsENB or srsUE. Although, during generation-based fuzzing (that uses mutated RRC ASN.1 schema), their application logs contain several errors like "\textit{Invalid field access}", "\textit{Buffer size limit was achieved}", and "\textit{Failed to unpack}". It is yet to be seen if running Berserker for more time (with higher stop values for seeds) will uncover any vulnerability in srsENB and srsUE.

\textbf{Uplink RRC and NAS messages coverage.} There are in total 31 types of uplink RRC messages in 3GPP TS 36.331 version 15.4.0 as listed in Table \ref{tab:rrc_messages_in_3gpp}. In mutation-based fuzzing, only the first six uplink messages (shown earlier in Figure \ref{fig:rrc_sequence}) are covered. In generation-based fuzzing, all 31 messages are covered. Regarding NAS message coverage, even though only four uplink messages (shown earlier in Figure \ref{fig:rrc_sequence}) are actually present in an intercepted RRC, they are treated as a blob and mutated. Thus, they will be either garbage or one of 18 partially valid uplink NAS messages defined in 3GPP TS 24.301 \cite{3gpp_eps_nas}.

\textbf{Downlink RRC and NAS messages coverage.} Similarly, there are in total 18 types of downlink RRC messages -- excluding broadcast messages -- as listed in Table \ref{tab:rrc_messages_in_3gpp}. While only the first six messages are covered in mutation-based fuzzing, all 18 are covered in generation-based fuzzing. Regarding NAS, even though only four messages (Figure \ref{fig:rrc_sequence}) are actually present in an intercepted RRC, they are treated as a blob and mutated. Thus, they will be either garbage or one of 18 partially valid downlink NAS messages.

\begin{figure}[ht!]
	\centering
	\begin{minipage}[]{0.48\linewidth}
		\includegraphics[width=\linewidth]{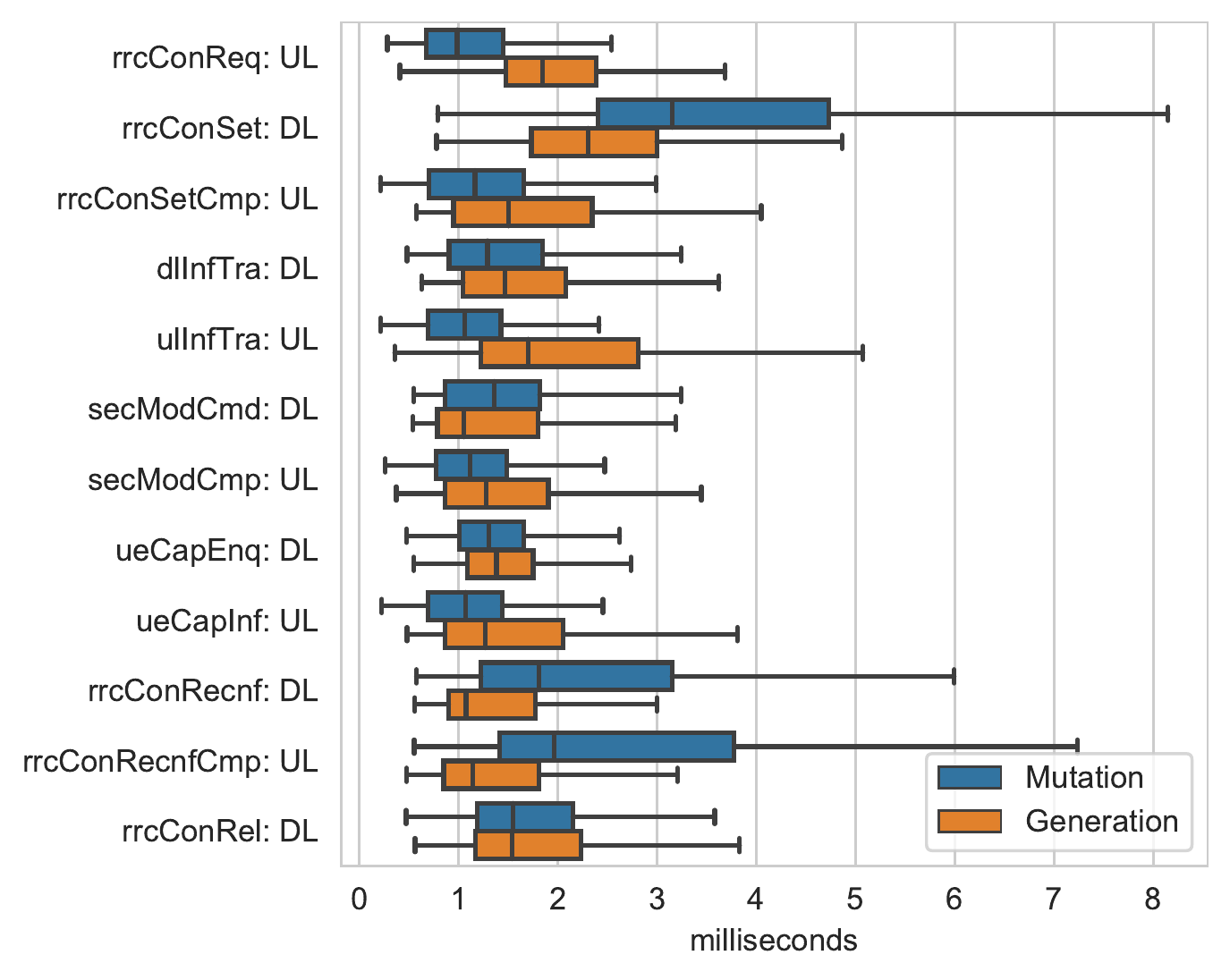}
		\caption {Box plot of the Fuzzer time cost, i.e., time taken between receiving a RRC PDU and producing a RRC PDU* in our experiments. Outliers and radamsa are omitted.} 
		\label{fig:latency_rrc_pdu_star} 
	\end{minipage}
	\quad
	\begin{minipage}[]{0.48\linewidth}
		\centering
		\includegraphics[width=\linewidth]{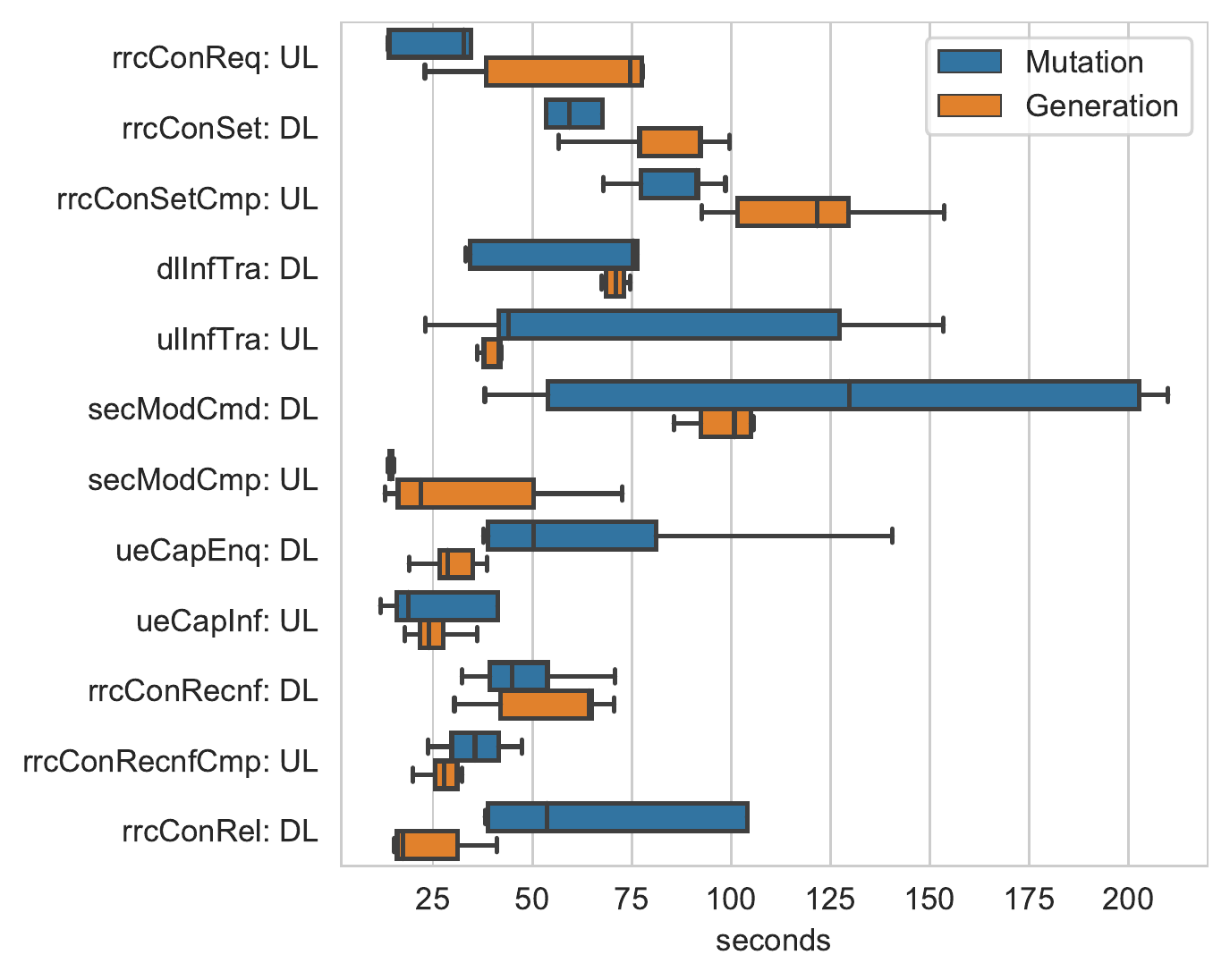}
		\caption {Box plot of the Driver time cost, i.e., average time taken in total to fuzz a RRC PDU in our experiments. Outliers and radamsa are omitted.} 
		\label{fig:latency_rrc_pdu_star_all} 
	\end{minipage}
\end{figure}

\textbf{Time cost.}  We analyzed time costs introduced by the Fuzzer and Driver components of Berserker, corresponding to our experimental results. 

The Fuzzer time cost, shown in Figure \ref{fig:latency_rrc_pdu_star}, is the time taken between receiving a RRC PDU and producing a RRC PDU*. With this time cost, we can assess whether Berserker is within the processing delay requirements for RRC procedures specified by 3GPP.

For uplink fuzzing, delay requirements on the UE side are relevant, which are specified in Clause 11.2 of 3GPP TS 36.331 for 4G:
\begin{itemize}[noitemsep,topsep=2pt,parsep=0pt,partopsep=0pt]	
	\item RRCConnectionRequest: n/a because it is initiated by the UE.
	\item RRCConnectionSetupComplete: within 15ms.
	\item SecurityModeComplete: within 10ms.
	\item UECapabilityInformation: within 10 or 80ms.
	\item ULInformationTransfer: n/a because it is initiated by the UE.
	\item RRCConnectionReconfigurationComplete: within 20ms.	
\end{itemize}

Similar requirements for 5G are defined in Clause 12 of 3GPP TS 38.331, which are more stringent with minimum requirements up to 5ms. As seen in Figure \ref{fig:latency_rrc_pdu_star}, Berserker mostly introduces 1-2ms of delay for uplink fuzzing which is well within the range of 4G and 5G requirements. In the uplink, there is also a pattern that generation-based fuzzing mostly has higher Fuzzer time cost than mutation-based. This can be attributed to the fact that generation-based fuzzing involves traversing RRC message structures and producing several OPTIONAL fields.

For downlink fuzzing, delay requirements on the RAN side are relevant. But 3GPP RRC TSes do not specify delay requirements on the RAN side, which could be explained by the fact that RAN itself is responsible for scheduling. Nevertheless, as seen in Figure \ref{fig:latency_rrc_pdu_star}, Berserker introduces 1-3ms of delay for downlink fuzzing, which is in similar range as for uplink. 

The Driver time cost, shown in Figure \ref{fig:latency_rrc_pdu_star_all}, is the average time taken for each strategy of fuzzed messages between starting srsLTE applications and stopping them after last IP ping request. This time cost helps estimating how long it will take in total to run experiments with a setup like Berserker's. The figure shows that time taken to fuzz one RRC message could be between 20s to 125s which are not low numbers. This comes from the non-aggressive implementation of the Driver that prioritizes stability over speed, e.g., setting inactivity timer in srsENB to 5s; backing off for 10s after each successful IP ping and 20s after each failed IP ping. These lead to long testing times on average, especially when fuzzed messages hinder connection establishment causing multiple failed IP pings (largely with RRC messages with a tunneled NAS message).


\textbf{Limitations.} Berserker detects vulnerabilities solely based on logs produced by SUTs' applications. Therefore, it will miss the problems that are not directly visible in logs, e.g., increasing memory consumption. Ability to probe the SUTs' system resources like CPU and memory usage would be a useful addition to Berserker.

It also means that Berserker treats the SUTs as blackboxes and does not comprise of instrumentation of the SUTs' binaries. Therefore, it lacks quantitative code coverage. Berserker needs to be upgraded to do coverage-guided fuzzing or even integrated with symbolic executions. Coverage-guided fuzzing coupled with feedback from SUT could produce meaningful testcases as proposed by \cite{Song2019SPFuzzAH}. 

Although Berserker sends out-of-sequence messages in generation-based fuzzing, it does so in random without any knowledge of RRC or NAS protocol states. Berserker could be enhanced to do smart sequence fuzzing.

Currently, NAS messages are treated as random blob. Adding knowledge of NAS protocol (e.g., by integrating T-Fuzz \cite{Johansson2014TFuzzMF}) to Berserker would enhance its ability to find more vulnerabilities. 

The Driver part of Berserker could be improved in terms of speed, e.g., with shorter back offs, simultaneous iterations with multiple instances of concrete UE and RAN/CN implementations, and SUT applications.

\section{Conclusion} \label{conclusions}
Berserker's ASN.1-based design shows good results by discovering previously unknown vulnerabilities in two open-source telecom projects, srsLTE and openLTE. It covers fuzzing of not only the RRC protocol but also the NAS protocol tunneled in RRC. In fact, the discovered vulnerabilities are caused by improper parsing and security processing of NAS messages by the network. Discovery of vulnerabilities provides manufacturers of mobile phones and network equipments an opportunity to fix them. Hence, Berserker is very relevant to increasing telecom networks' robustness. Its main advantage comes from requiring minimal maintenance after initial integration with a system under test, because it is compatible with any version of 4G and 5G 3GPP RRC technical specifications. It extracts RRC ASN.1 schema directly from these specifications and can additionally sidestep the constraints on sizes and types imposed by them. Berserker relies on concrete implementations of telecom protocol stack and therefore is unaffected by lower layer protocol handlings. There are opportunities to further improve Berserker, e.g., in terms of coverage-guided fuzzing, sequence fuzzing, and observability.

\bibliographystyle{unsrt}
\bibliography{references}


\end{document}